\documentclass[journal,twoside]{IEEEtran}

\newtheorem{myDef}{Definition}
\usepackage{cite}
\usepackage{amsmath,amssymb,amsfonts}
\usepackage{bm}
\usepackage{algorithmic}
\usepackage{graphicx}
\usepackage{epstopdf}
\usepackage{textcomp}
\usepackage{xcolor}
\usepackage{stfloats}
\usepackage{subfigure}
\usepackage{diagbox}
\usepackage{enumerate} 
\usepackage{threeparttable}
\usepackage{multirow}
\usepackage{booktabs}
\usepackage{color}
\usepackage{url}
\usepackage[colorlinks,
linkcolor=black,
anchorcolor=black,
urlcolor=black,
citecolor=black]{hyperref}

\ifCLASSINFOpdf
\else
\fi
\begin{document}
%
\title{Detecting Mixing Services via Mining Bitcoin Transaction Network with Hybrid Motifs}
%
%
%

\author{Jiajing~Wu,~\IEEEmembership{Senior Member,~IEEE,}
	Jieli~Liu,
	Weili~Chen,
	Huawei~Huang,~\IEEEmembership{Member,~IEEE,}\\
	Zibin~Zheng,~\IEEEmembership{Senior Member,~IEEE,}
	and~Yan~Zhang,~\IEEEmembership{Fellow,~IEEE}
	\thanks{Manuscript received January 26, 2020; revised September 22, 2020; accepted December 29, 2020. The research is supported by the  Key-Area Research and Development Program of Guangdong Province (No. 2018B010109001), and the National Natural Science  Foundation of China (No. 62032025, No. 61973325, No.U1811462). (\textit{Corresponding Author: Zibin Zheng})}
	\thanks{J. Wu, J. Liu, W. Chen, H. Huang and Z. Zheng are with the School of Computer Science and Engineering, Sun Yat-sen University, Guangzhou 510006, China. (Email: wujiajing@mail.sysu.edu.cn, zhzibin@mail.sysu.edu.cn)}
	\thanks{Y. Zhang is with the Department of Informatics, University of Oslo, Oslo 0316, Norway.}
}

%
%

\markboth{IEEE Transactions on Systems, Man, and Cybernetics: Systems,~Vol.~XX, No.~XX, xxx~2020}%
{Wu \MakeLowercase{\textit{et al.}}: Detecting Mixing Services via Mining Bitcoin Transaction Network with Hybrid Motifs}
%



\maketitle

\begin{abstract}
As the first decentralized peer-to-peer (P2P) cryptocurrency system allowing people to trade with pseudonymous addresses, Bitcoin has become increasingly popular in recent years. However, the P2P and pseudonymous nature of Bitcoin make transactions on this platform very difficult to track, thus triggering the emergence of various illegal activities in the Bitcoin ecosystem. Particularly, \emph{mixing services} in Bitcoin, originally designed to enhance transaction anonymity, have been widely employed for money laundering to complicate the process of trailing illicit fund. In this paper, we focus on the detection of the addresses belonging to mixing services, which is an important task for anti-money laundering in Bitcoin. Specifically, we provide a feature-based network analysis framework to identify statistical properties of mixing services from three levels, namely, network level, account level and transaction level. To better characterize the transaction patterns of different types of addresses, we propose the concept of Attributed Temporal Heterogeneous motifs (ATH motifs). Moreover, to deal with the issue of imperfect labeling, we tackle the mixing detection task as a Positive and Unlabeled learning (PU learning) problem and build a detection model by leveraging the considered features. Experiments on real Bitcoin datasets demonstrate the effectiveness of our detection model and the importance of hybrid motifs including ATH motifs in mixing detection.
\end{abstract}


%
\IEEEpeerreviewmaketitle

\section{Introduction}
\IEEEPARstart{B}{itcoin}, the world's first peer-to-peer (P2P) cryptocurrency system\cite{nakamoto2008bitcoin}, has become one of the hottest buzzwords with a dominant share of the cryptocurrency market \cite{yuan2018blockchain} due to its pseudonymous nature in decentralized trading process as well as its low transaction fees.

\begin{figure}[htbp]
	\centerline{\includegraphics[scale=0.132,trim=20 20 20 0]{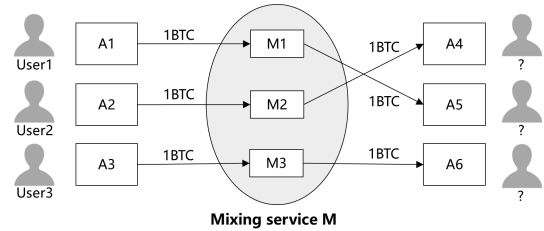}}
	\caption{An example of mixing services, which can conceal the identity of users and complicate fund tracing by participating in a transaction with multiple users.}
	\label{fig1}
\end{figure} 

However, the P2P and pseudonymous nature of Bitcoin make transactions on this platform very difficult to track, thus triggering the emergence of various illegal activities in the Bitcoin ecosystem \cite{chen2018blockchain}. For instance, about 7,000 Bitcoins which worth \$40.7 million were stolen from Binance recently \cite{3360664}, one of the largest cryptocurrency exchanges in the world. Then the stolen Bitcoins can be cashed out directly through exchanges.
However, before conducting the businesses, exchanges typically implement the Know-Your-Customer (KYC) process, which is widely adopted in traditional e-payment scenarios to verify the identities of the users, review their financial activities, and ascertain what risks they may pose. With the enforcement of the KYC process, the identity of the thieves can be easily exposed via the identity information provided by the exchanges, and the stolen Bitcoins usually need to be laundered into ``clean'' Bitcoins by some techniques before being cashed out. It has been demonstrated that, mixing services such as BitLaundry, Helix Light, Bitcoin Fog, etc., have involved in this process of {\em money laundering}~\cite{M2014An} and can be regarded as significant tools for concealing illicit profits in Bitcoin.

Bitcoin mixing services are originally designed to enhance the anonymity of transactions and make the sources of funds more untraceable. Fig.~\ref{fig1} gives a simple illustration of Bitcoin mixing. Three users represented as $A1$, $A2$ and $A3$ send 1 Bitcoin (abbreviation BTC) to three addresses $M1$, $M2$ and $M3$ of a mixing service $M$, respectively, and provide their own new addresses $A4$, $A5$, and $A6$ to receive the Bitcoin back. Then $M$ randomly select an address from $M1$, $M2$ and $M3$ to return money to $A4$, $A5$, and $A6$. In this way, the relationships between sources and destinations are confused, thus increasing the difficulty of tracing the source of funds and analyzing the transaction behavior of users. 
Since Bitcoin is designed with pseudonymous identities and the real identity behind an address can be learned only when the user use this address to buy or sell Bitcoins with an exchange, it is unlikely to enforce the KYC process for regulation. Therefore, the study on identification of mixing services and tracing illegal transactions in Bitcoin is of great value for building a healthier Bitcoin ecosystem.

\begin{figure}[htbp]
	\centerline{\includegraphics[scale=0.19,trim=20 10 20 0]{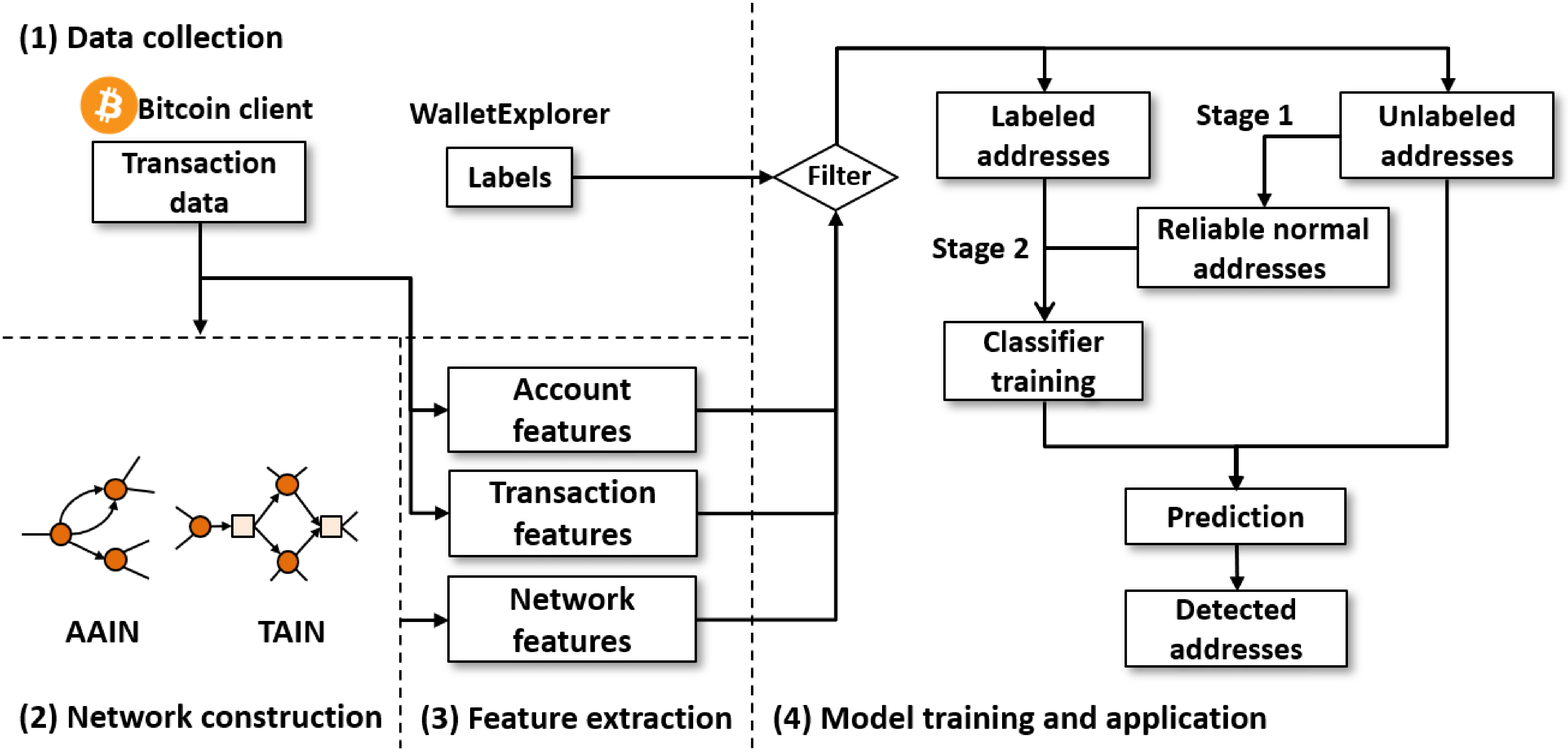}}
	\caption{An overview of  the proposed Bitcoin mixing detection framework including four modules, namely, data collection, network construction, feature extraction, model training and application.}
	\label{fig2}
\end{figure}

Fortunately, the public and irreversible transaction records provide us an opportunity to detect irregular transaction patterns in Bitcoin. To this end, in this paper, we focus on detecting addresses belonging to mixing services via mining the transaction records and attempt to characterize their transaction patterns. Based on the detection results, we can further chase up users involved in criminal activities by analyzing users who take part in Bitcoin mixing.

In recent years, several studies have shed light on the problem of detecting Bitcoin mixing services. It has been reported that mixing services and exchanges are two key components in laundering Bitcoins \cite{M2014An,van2018bitcoin} while mixing services have a higher propensity to be used in laundering illicit money \cite{fanusie2018bitcoin}. To answer how mixing services work, the operation model of several mixing services were studied by reverse-engineering methods in \cite{M2014An}. Based on the observations given by \cite{M2014An}, Prado-Romero et al. proposed the problem of mixing detection and tackled this problem by exploiting the method of community outlier detection \cite{Prado2017Discovering}. Yet till now, Bitcoin mixing detection is still an extremely tricky task due to several great challenges as follows: (1) \textbf{Incomplete label information.} Labeled addresses associated with mixing services occupy only a small fraction of all addresses, and the true identities of most other addresses are unknown in Bitcoin. (2) \textbf{Dynamic process with multiple transactions.} Some mixing services use hubs to combine multiple transactions or split a large amount of money into multiple smaller transactions, thus making it more difficult to identify the mixing processes as well as the addresses involved in the processes.
(3)\textbf{Various obfuscation patterns.} Mixing services are provided by different third-party platforms, and their obfuscation patterns vary a lot from each other.

In this work, to deal with the problem of incomplete label information, we tackle the task of Bitcoin mixing detection as a Positive and Unlabeled learning (PU learning) problem \cite{liu2003building} and then adopt a two-stage strategy to enhance the detecting performance.
In order to analyze the transaction records more comprehensively, we construct two kinds of temporal directed transaction networks including a homogeneous Address-Address Interaction Network (AAIN) and a heterogeneous Transaction-Address Interaction Network (TAIN), to depict the relationship between addresses and the relationship between addresses and transactions, respectively.  
Network motifs have been widely proven to be a powerful tool in handling various network mining tasks \cite{milo2002network, paranjape2017motifs, li2018community}. 
To better analyze the complicated dynamic processes in the Bitcoin transaction network, we propose a novel concept called \textbf{ Attributed Temporal Heterogeneous motifs (ATH motifs)} for the TAIN. The hybrid motifs composed of temporal homogeneous motifs in AAIN and ATH motifs in TAIN, are employed as the vital features for the detection of mixing services. 

As shown in Fig.~\ref{fig2}, the proposed mixing detection framework mainly contains four modules: (1) \textbf{Data collection}, which gathers the Bitcoin transaction data from a Bitcoin client and crawls the label information from WalletExplorer\footnote{https://www.walletexplorer.com.}. (2) \textbf{Network construction}, constructing AAIN and TAIN from the transaction records for feature extraction. (3) \textbf{Feature extraction}, whose purpose is to extract features from multiple levels. (4) \textbf{Model training and application}, which trains the model using the training set, makes prediction for the unlabeled addresses and finally outputs the detected mixing addresses. 

In summary, the main contributions of this paper can be listed as follows:

(1) To the best of our knowledge, we are the first to apply network motifs on the problem of Bitcoin mixing detection. We propose the novel concept of ATH motifs and demonstrate that both temporal and ATH motifs play an important role in Bitcoin mixing detection.  

(2) We propose a feature-based transaction network analysis framework and generalize the issue of Bitcoin mixing detection as a PU learning problem, the purpose being to make better use of the labeled addresses under the precondition of imperfectly labeled datasets.

(3) The proposed model achieves a high true positive rate and a low false positive rate in Bitcoin mixing detection, which facilitates fund tracing and crime detection in the Bitcoin ecosystem.

The remaining sections of this paper are organized as follows. Sections~\ref{DC} to \ref{DM} introduce the details of the aforementioned four modules of the proposed mixing detection framework one by one. Then we present experimental results in Section \ref{ER}. Finally, we provide some related work in Section \ref{RW} and conclude this paper in Section \ref{C}.

\section{Data Collection}\label{DC}

%

\begin{table}
	\renewcommand{\arraystretch}{1.15}
	\begin{threeparttable}[b]
		\caption{\label{table1}Statistics of the datasets.}	
		\centering	
		\setlength{\tabcolsep}{0.4mm}{	
			\begin{tabular}{c|c|c|c|c|c}		
				\hline		
				\multirow{2}*{Dataset} & \multirow{2}*{Start time} & \multirow{2}*{Unlabeled address} & \multicolumn{3}{c}{Labeled address$^1$}\\
				\cline{4-6}	
				&&& BitcoinFog & BitLaunder & HelixMixer\\		
				\hline
				\hline
				2014&00:00, Nov. 1&2,507,872&6088&8&0\\
				\hline
				2015&00:00, Jun. 1&2,525,038&3911&9&2\\
				\hline
				2016&00:00, Jan. 1&2,502,738&198&2&3856\\
				\hline
		\end{tabular}}
		\begin{tablenotes}
			\item[1] Addresses of three mixing services including Bitcoin Fog, BitLaunder and Helix Mixer crawled from WalletExplorer are as our labeled addresses.
		\end{tablenotes}
	\end{threeparttable}
\end{table}

\begin{figure}[h]
	\centering
	\subfigure[A $3$-node, $3$-edge, $\delta$-temporal motif $M$.]{
		\label{fig3a}
		\includegraphics[scale=0.13, trim = -200 50 -200 30]{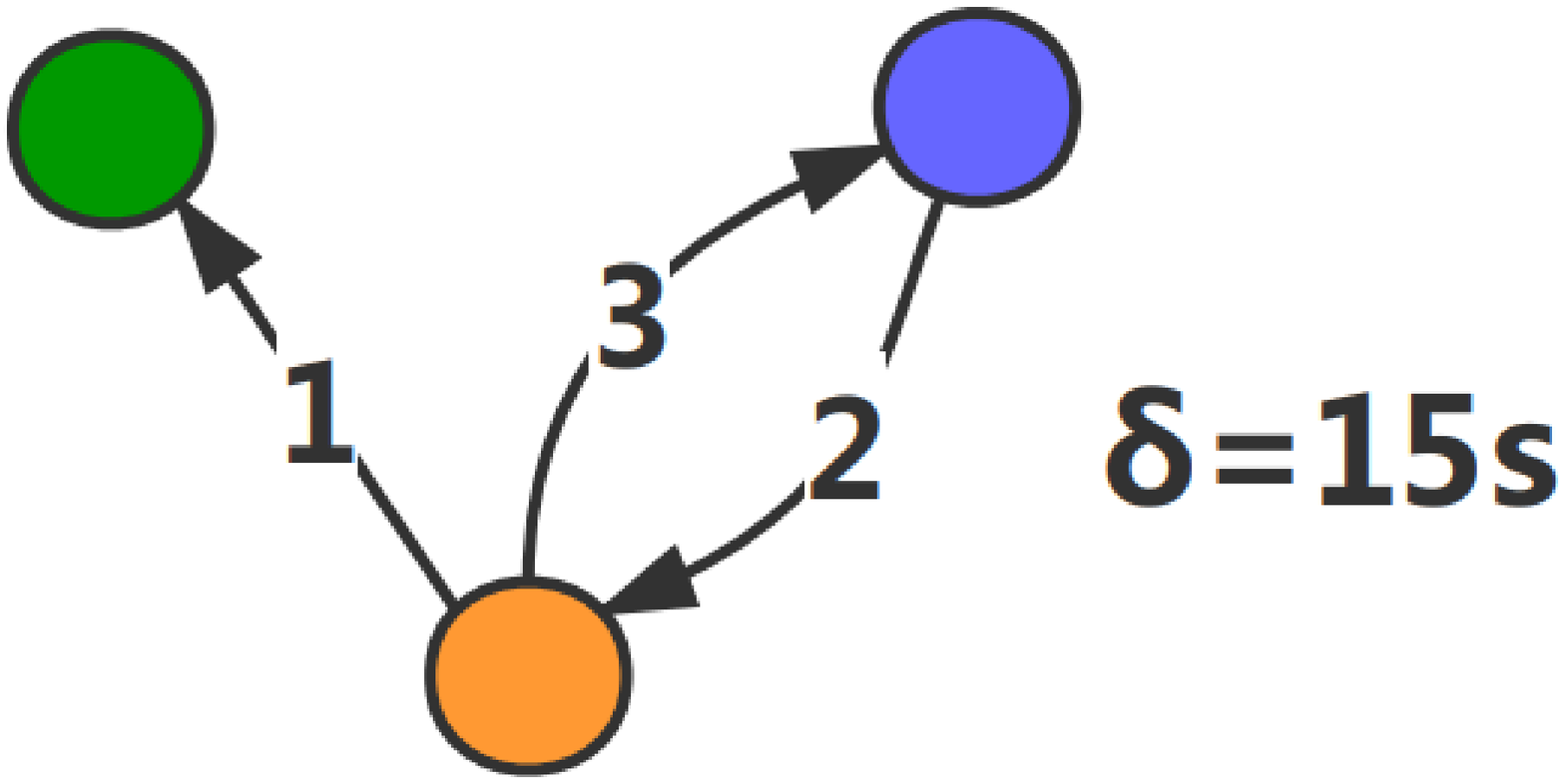}}  
	
	\subfigure[Two motif instances of $M$.]{
		\label{fig3b}
		\includegraphics[scale=0.13, trim=0 40 0 50]{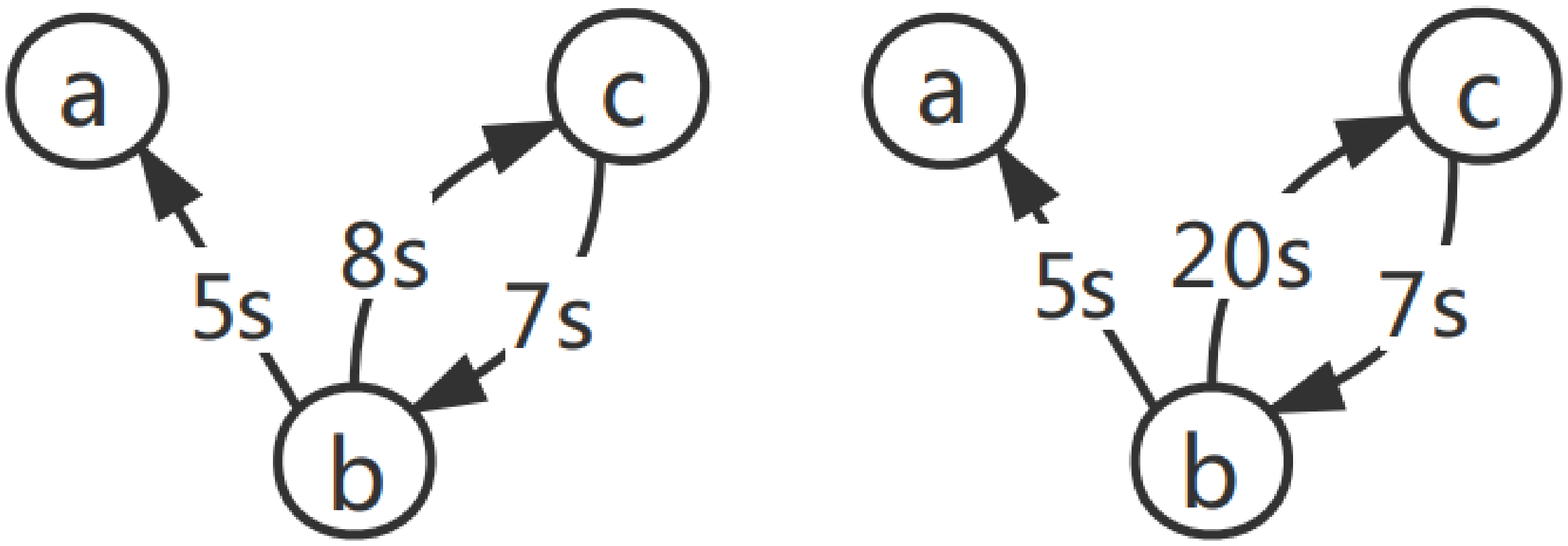}}
	\subfigure[Two graph patterns which are not instances of $M$.]{	
		\label{fig3c}
		\includegraphics[scale=0.13, trim= 0 35 0 50]{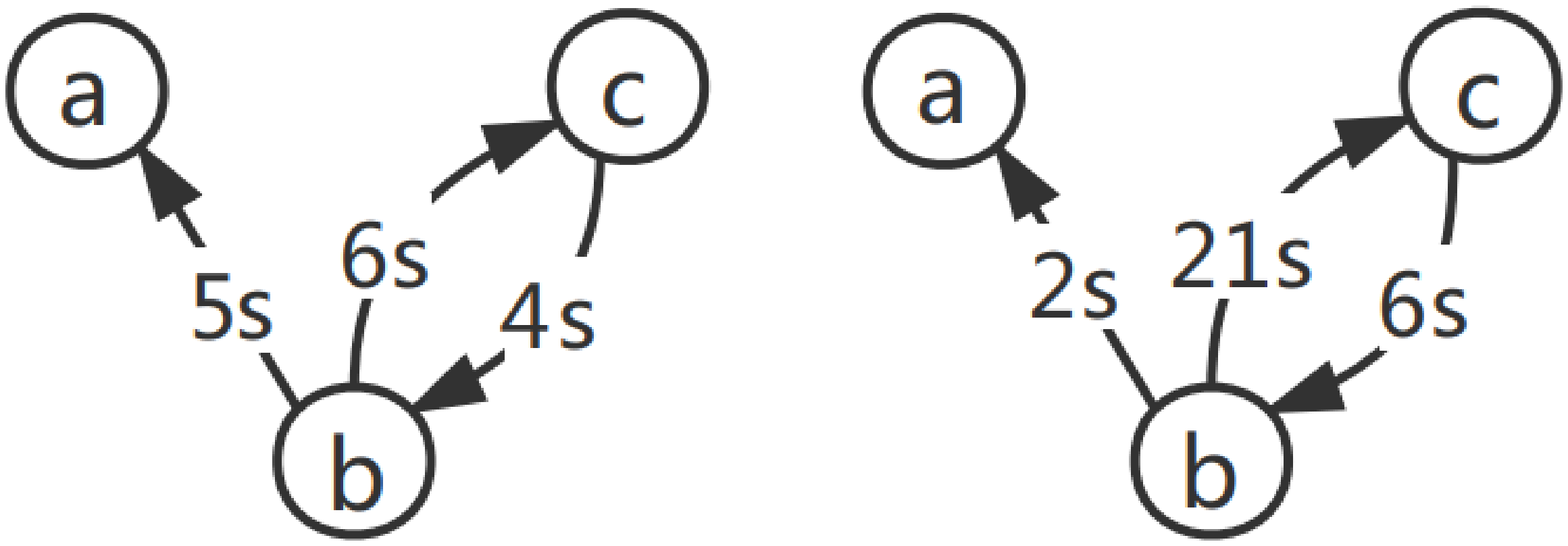}}
	\caption{ An example of a motif (a) and its instances (b). Graph patterns in (c) are not instances of $M$ because of their out of order edge sequence or their out of range edge occurring time (the constrained time window $\delta$ is set as 15s).}
	\label{fig3}
\end{figure}

WalletExplorer is a smart Bitcoin block explorer providing label information of addresses by making transactions with some services and observing how the Bitcoin flows merge. However, the name database of WalletExplorer no longer updated since 2016, which means that WalletExplorer does not include the new emerging services. The transaction data of Bitcoin are contained in blocks orderly, and they are publicly accessible by running a Bitcoin client. Considering the sufficiency of labeled addresses for training and the huge volume of Bitcoin transaction records, we extract three snapshots of Bitcoin transaction data between November 2014 and January 2016 with six months being the sampling interval. Each snapshot contains 1,500,000 transaction records. And we crawl the labeled addresses belonging to mixing services from WalletExplorer. The three snapshots with label information are referred to as the 2014, 2015 and 2016 datasets.

The labeled addresses obtained from WalletExplorer are mainly belonging to three mixing services as follows:
\begin{itemize}
	\item Bitcoin Fog{\footnote{https://bitcointalk.org/index.php?topic=50037.0}} is one of the earliest mixing services and can be accessed only via Tor. When using this service, each withdrawal will be split into a random number of transactions spreading out randomly over a specific time period.
	\item BitLaunder announced that it was ``the best Bitcoin laundry and Bitcoin laundering service''. However, it has been reported as one of the weakest mixers of all tested in the analysis conducted by Balthasar and Hernandez-Castro \cite{BalthasarH17}. Unfortunately, the detailed information of this service is not available anymore.
	\item Helix (has been offline since 2017{\footnote{https://bitcointalk.org/index.php?topic=5238537.0}}) offered two version of mixing services, including a standard version and a light version. The standard version required their users to register a wallet, and then the Bitcoins sent to the wallet would be automatically mixed and finally sent to a defined address. While for the light version, the Bitcoins could be withdrawn to up to five addresses. 
\end{itemize}

Table \ref{table1} shows the statistics of these three datasets. On the average, the labeled addresses only account for about 0.19\% of all addresses appearing in the transaction data.

\section{Network Construction and Motif Definition}\label{NC}
\begin{figure}[h]
	\centering
	\subfigure[]{
		\label{fig4a}
		\includegraphics[scale=0.14,trim=20 25 -30 0]{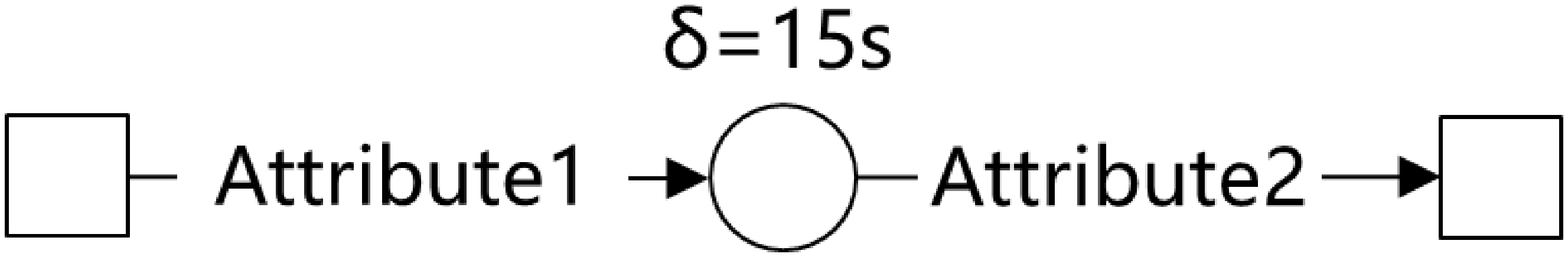}}
	
	\subfigure[]{	
		\label{fig4b}
		\includegraphics[scale=0.14,trim=0 20 -150 20]{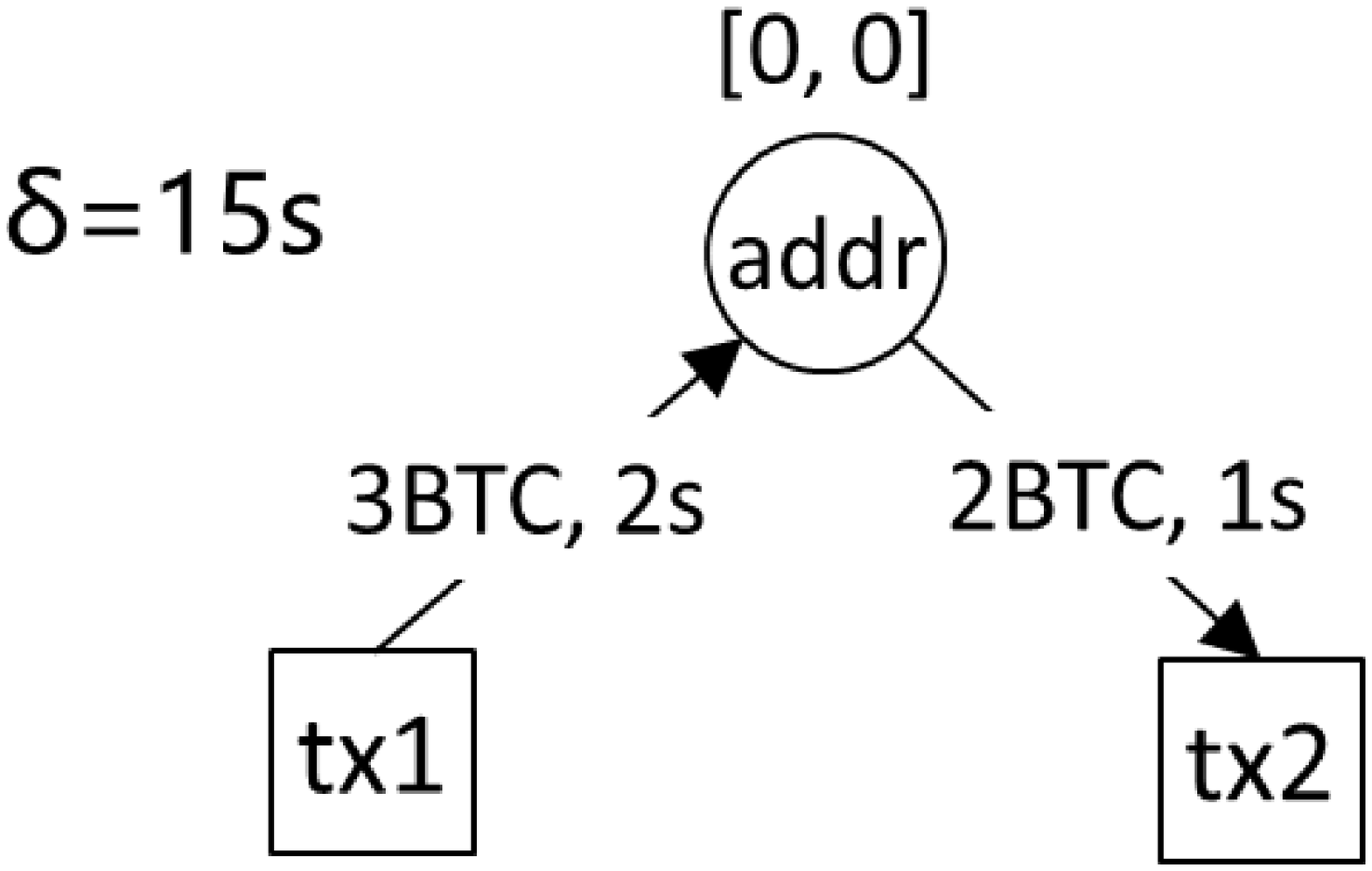}}
	\subfigure[]{	
		\label{fig4c}
		\includegraphics[scale=0.14,trim=0 20 -150 20]{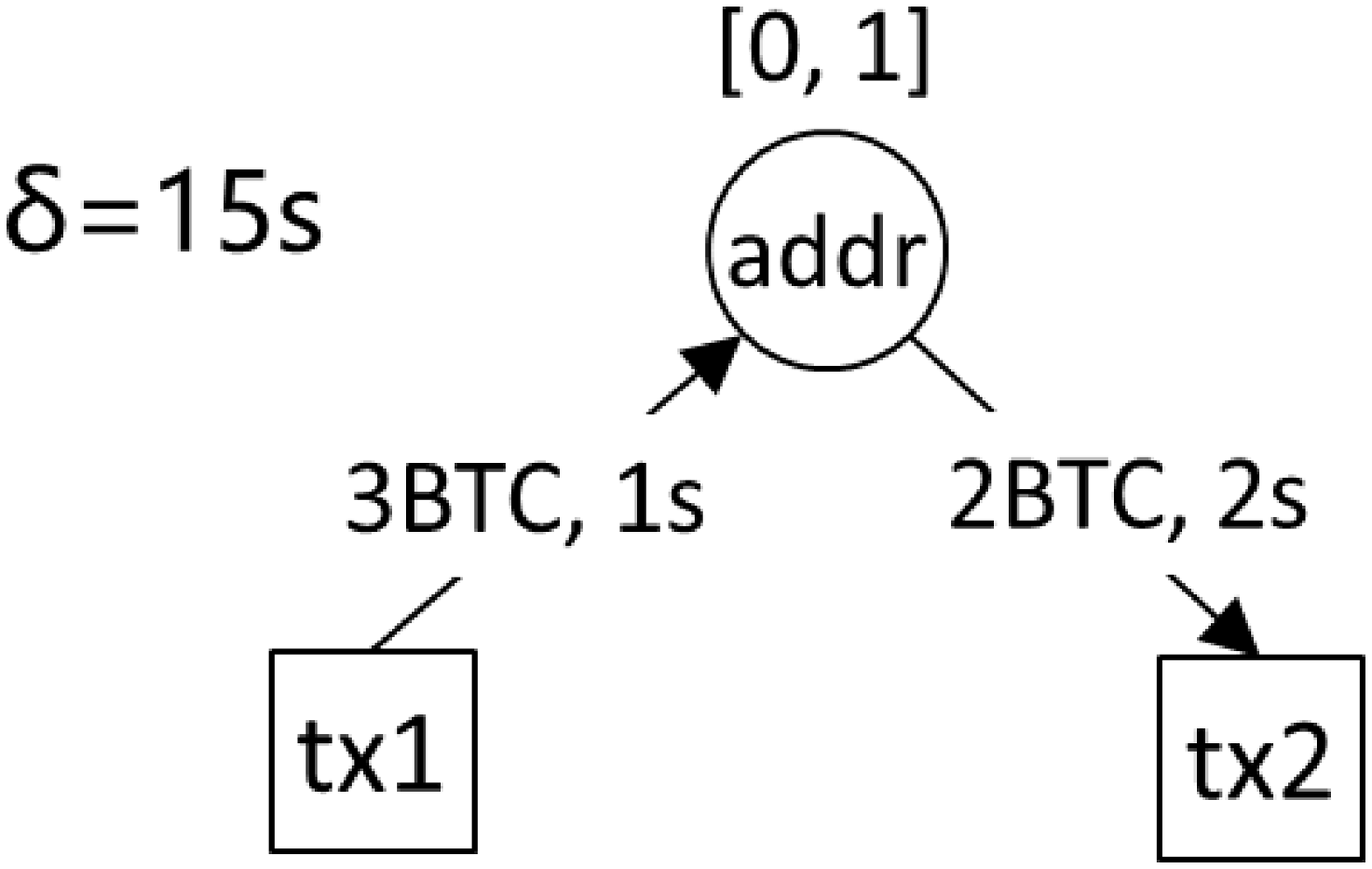}}
	
	\subfigure[]{	
		\label{fig4d}
		\includegraphics[scale=0.14,trim=0 20 -150 20]{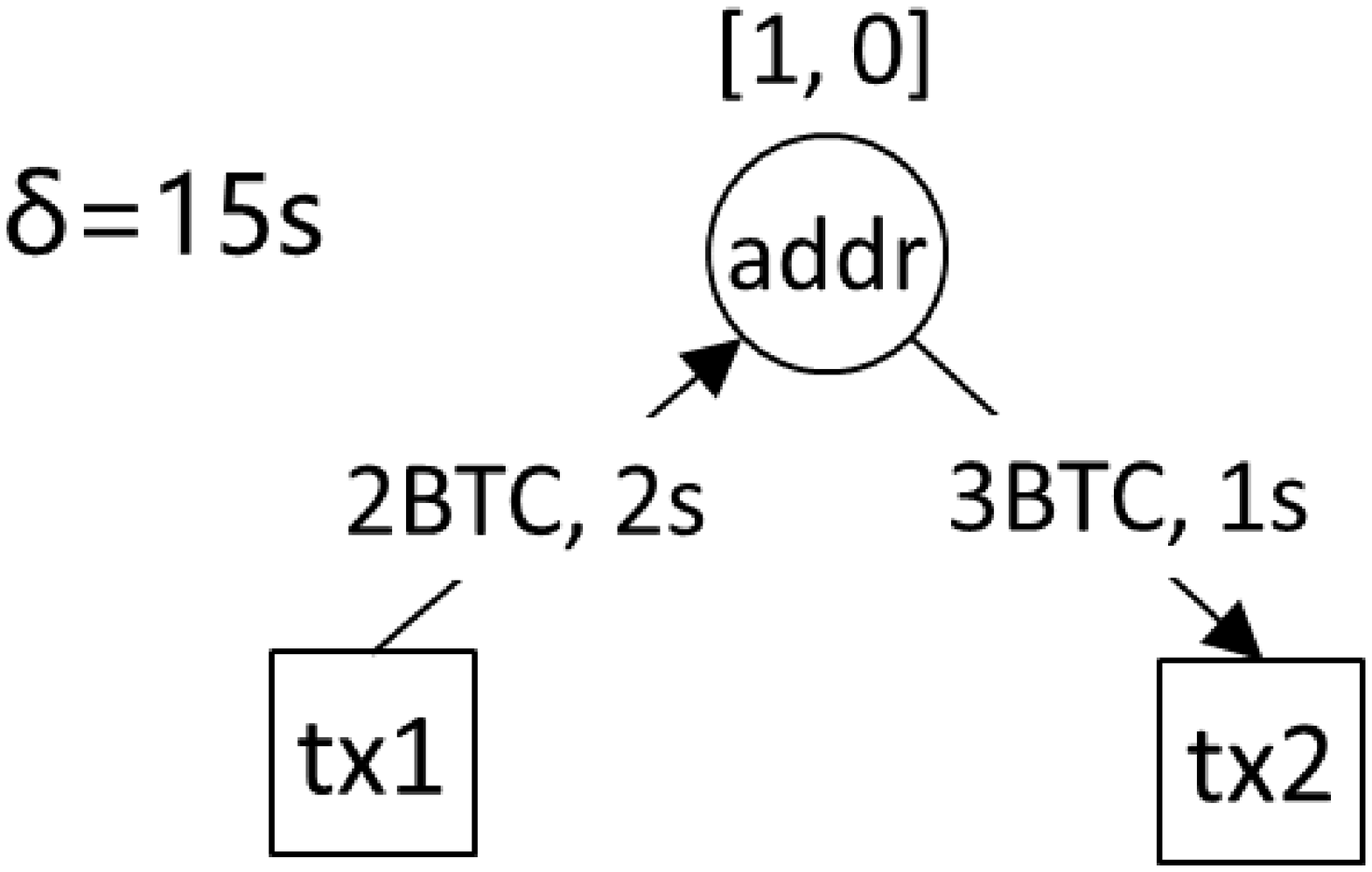}}
	\subfigure[]{	
		\label{fig4e}
		\includegraphics[scale=0.14,trim=0 20 -150 20]{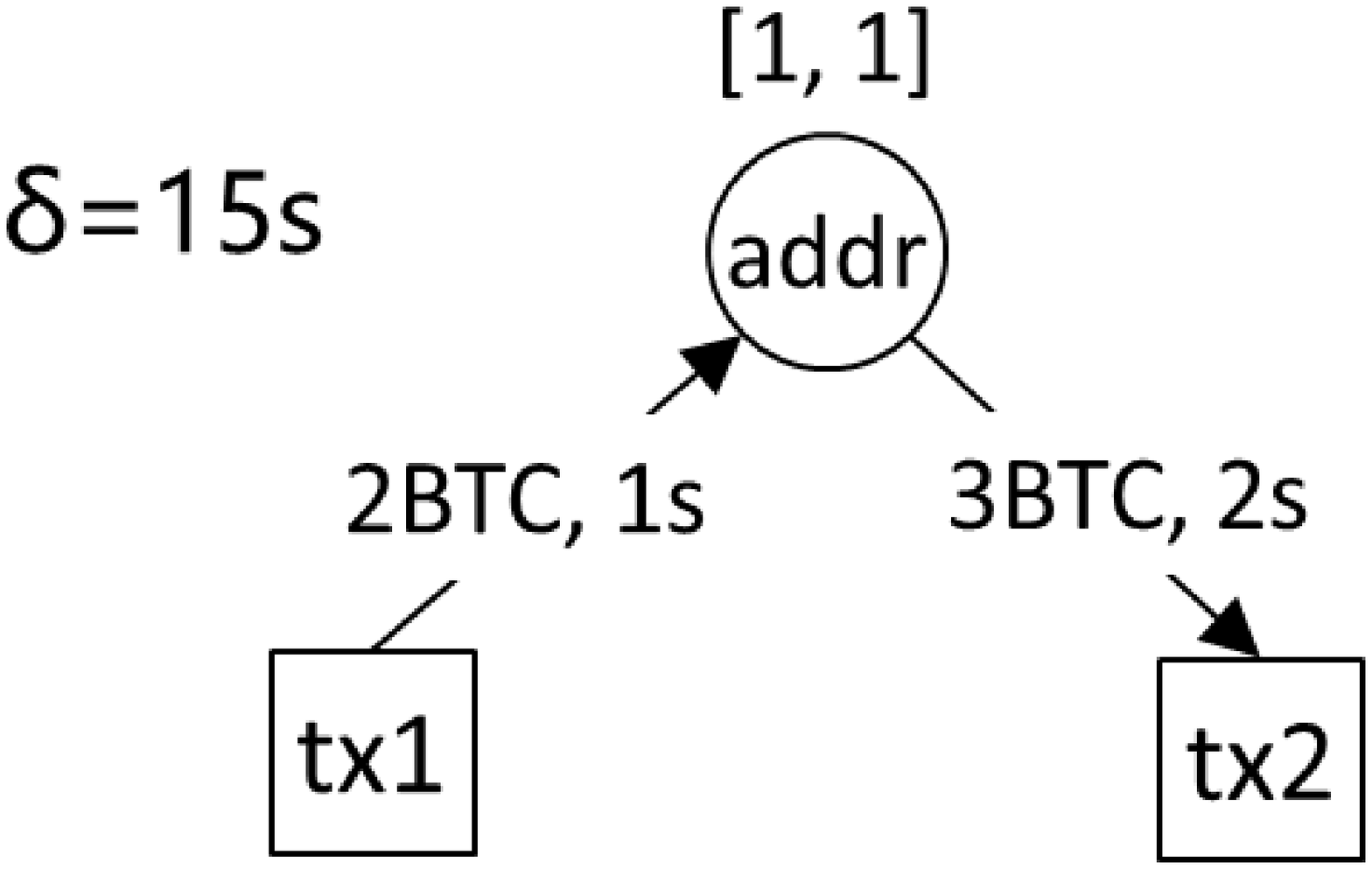}}
	\caption{(b)-(e) are four instances of the ATH motif (a) in transaction network. The edge vector ${\varGamma}_V$, which maps the amount value attribute to the first bit and the time information to the second bit, can differentiate varieties of transaction patterns with the same topology. In this case, the first bit of ${\varGamma}_V$ is set as 0 if the amount of $tx1$ is higher than $tx2$, and as 1 otherwise; the second bit of ${\varGamma}_V$ is set as 0 if the time of $tx1$ is later than $tx2$, and as 1 otherwise.}
	\label{fig4}
\end{figure}

Transaction records of Bitcoin can be abstracted as a huge complex network, where each node refers to a Bitcoin address and each edge represents a transaction process between addresses. Using this simple modeling method, we construct a homogeneous Address-Address Interaction Network (AAIN) to investigate the interaction patterns of addresses. Since Bitcoin transactions usually involve multiple inputs and multiple outputs, from this figure depicting the transaction relationships between address pairs, it is difficult to figure out how much an address has taken from another address. To this end, we construct a heterogeneous Transaction-Address Interaction Network (TAIN) to represent the transaction amount information. This is an attributed temporal heterogeneous information network (HIN) where a node can be either a particular transaction or an address. From TAIN, we can clearly find how much an address has sent to or received from a transaction. Therefore, compared with AAIN, TAIN can display the strength of money transfer more clearly.
Network motifs, which can be regarded as the recurrent small subgraph patterns in networks, have been demonstrated as an important tool for characterizing higher-order interactions and understanding various properties of complex systems \cite{milo2002network,paranjape2017motifs}.
In the following, we present the definition of AAIN, TAIN and their motifs in detail. 

\subsection{AAIN and Temporal Motifs}
\begin{myDef}[AAIN] An Address-Address Interaction Network (AAIN) is a temporal direct multigraph $G=(V,E)$, where $V$ is the set of nodes and $E$ is the set of edges carrying temporal information. Each node $v \in V$ denotes a Bitcoin address and each edge $e \in E$ standing for a transaction is defined as a tuple $(u,v,tx,t)$, denoting that address $u$ is a source and address $v$ is a destination for a transaction $tx$ happening at time $t$. 
\end{myDef}


\begin{myDef}[Temporal Motifs]
	Temporal motifs are defined as recurring interconnection patterns occurring in temporal networks \cite{paranjape2017motifs}. Particularly, a $k$-node, $l$-edge, $\delta$-temporal motif instance $M^{k,l}_{\delta}(G)$ of a temporal network $G=(V,E)$ can be represented as
	\begin{center}
		$M^{k,l}_{\delta}(G)=(V^k_M,E^l_M,\delta)$,
	\end{center}
	where $V^k_M$ ($V^k_M\subseteq V$) is a set of $k$ nodes, $E^l_M$ ($E^l_M\subseteq E$) is a set of $l$ edges and $\delta$ is a time window indicating that all of edges in the motif occur within a $\delta$ duration, i.e., an increased sequence ${t_1, t_2, ..., t_l}$ which records the timestamp of each edge in the motif instance satisfies $t_1 \leqslant t_2 \leqslant ...\leqslant t_l$ and $t_l - t_1 \leqslant \delta$.  
\end{myDef}

Different from static network motifs, temporal motifs well preserve the time-ordered sequence of contacts in a time window, being effective in analyzing temporal structure of complex networks. Figs.~\ref{fig3}(a)-(b) illustrate a $3$-node, $3$-edge, $\delta$-temporal motif $M$ and its instances, while graph patterns in Fig.~\ref{fig3}(c) are not instances of $M$ because their edge order or occurring time window does not satisfy the condition.

\subsection{TAIN and ATH Motifs}

\begin{myDef}[TAIN]
	A Transaction-Address Interaction Network (TAIN) is an attributed temporal heterogeneous information network $G=(V,E,\varOmega)$ with ${\varphi}_V: V\rightarrow \{$address$, $transaction$\}$ for node type mapping, ${\varphi}_E: E\rightarrow \{\emph{transaction-address}, \emph{address-transaction}\}$ for edge type mapping, and $\varOmega$ denoting the set of attributes attached to edges in the graph including transaction amount and transaction time. A \emph{transaction-address} edge $(u,tx_{in},a,t_1)$ denotes that an input transaction $tx_{in}$ happens at time $t_1$ and transfers $a$ Bitcoins into an address $u$, while an \emph{address-transaction} edge $(v,tx_{out},b,t_2)$ denotes that an output transaction $tx_{out}$ happens at time $t_2$ and transfers $b$ Bitcoins out of an address $v$.
\end{myDef}


\begin{myDef}[Attributed Temporal Heterogeneous (ATH) Motifs]
	ATH motifs are local recurring subgraphs of attributed temporal HINs, described by a set of nodes, a set of edges, attributes and a time window. A $\delta$-ATH motif instance of an attributed temporal HIN $G=(V,E,\varOmega)$ can be defined as:
	\begin{center}
		$M^{\delta}_{ATH}(G) = (V_{ATH}, E_{ATH}, {\varGamma}_{\varOmega}, \delta)$,
	\end{center}
	where $V_{ATH}$ ($V_{ATH} \subseteq V$) represents the set of nodes, $E_{ATH}$ ($E_{ATH} \subseteq E$) represents the set of edges, also satisfying node types $|\{{\varphi}_{V}(v)| v \in V_{ATH}\}|>1$ or edge types $|\{{\varphi}_{E}(e)| e \in E_{ATH}\}|>1$. ${\varGamma}_{\varOmega}$ denotes a mapped vector of edge attributes, and $\delta$ is a time window that constrains $min(\psi_E(e)) + \delta \leq max(\psi_E(e))$ for $e \in E_{ATH}$, where $\psi_E$ is a time mapping which maps each edge $e \in E$ to its occurring time.
\end{myDef}

Though some subgraphs in the TAIN may share the same topology, the attribute information can make them different. Taking the ATH motif shown in Fig.~\ref{fig4} and its four instances as an example, the instance in Fig.~\ref{fig4c} represents the transaction pattern of receiving money first and sending out less money later, while the instance in Fig.~\ref{fig4d} stands for sending money out firstly and receiving less money later, which has an opposite transaction order and leads to a negative balance.

\section{Feature-based Analysis}\label{FE}

Due to the specific function of mixing Bitcoins, addresses associated with mixing services may have several unique features different from normal addresses. In the following, we aim to extract features of the addresses from three levels and conduct descriptive statistics on them.

\begin{figure}[h]
	\centering
	\subfigure[Candidate 2-edge $\delta$-temporal motif patterns in AAIN.]{
		\label{motifa}
		\includegraphics[scale=0.135,trim=-80 40 -80 50]{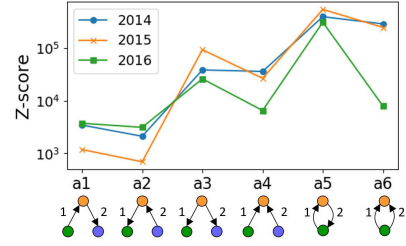}}
	\subfigure[Candidate $\delta$-ATH motif patterns in TAIN.]{
		\label{motifb}
		\includegraphics[scale=0.135,trim=0 30 0 50]{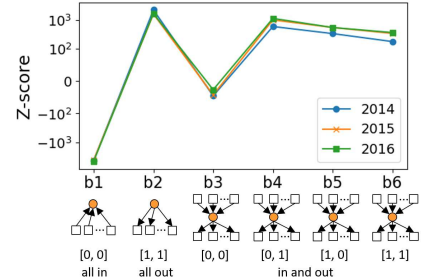}}
	\caption{The z-score value of the candidate motif patterns.}
	\label{z-score}
\end{figure}

\subsection{Network Features}

Different types of objects have different interaction ways in complex systems, which would affect the topological structure of the whole network. In this part, we extract network features from both AAIN and TAIN. To characterize the interaction patterns and reveal the functional properties in the network, we propose to take some higher-order network features (i.e. network motifs) into account. 

Network motifs are small subgraph patterns in a network that occurring with significantly higher frequency than those in randomized networks \cite{milo2002network}. The statistical significance of a pattern can be measured by z-score, calculated as:
\begin{equation}
{\rm z{-}score}=\dfrac{n_{real}-\bar{n}_{random}}{std(n_{random})},
\end{equation}
where $n_{real}$ denotes the frequency of the pattern occurring in the real network, $\bar{n}_{random}$ and $std(n_{random})$ denote the mean and variance of the pattern occurring frequencies in a set of random networks. A pattern is usually regarded as a statistically significant motif if its z-score $>$ 2.0 \cite{wong2012biological}.

We consider six simplest transaction patterns (Fig. \ref{motifa})  with two edges in AAIN, which illustrate how an address interacts with other addresses for a $\delta$ duration. For example, pattern $a1$ represents that an address first receives money from a neighbor and then transfers money to another neighbor while pattern $a2$ represents an opposite transaction order. Moreover, we abstract the transactions of an address occurring within a $\delta$ time window as three kinds of topological structures in TAIN: \emph{all in}, \emph{all out}, \emph{in and out}, which illustrate only input transactions, only output transactions, both input and output transactions occurring within the time window, respectively. By taking the amount information and temporal information into account, the direction and strength of Bitcoin transfer can be better reflected in these substructures. In our scenario, the relative size between the attribute information of input and output transactions is more important than their actual absolute value. Therefore, the \emph{in and out} structure can be further divided into four patterns, and all these six patterns (Fig. \ref{motifb}) can be represented as candidate $\delta$-ATH motifs with a binary value function defining each bit of  ${\varGamma}_{\varOmega}$ as:
\begin{equation}
\setlength\abovedisplayskip{10pt}
\setlength\belowdisplayskip{10pt}
{\varGamma}_{\varOmega}[0]=
\begin{cases}
0& \bar{v}_{in} \geq \bar{v}_{out}\\
1& \bar{v}_{in} < \bar{v}_{out}
\end{cases}, 
\setlength\abovedisplayskip{10pt}
\setlength\belowdisplayskip{10pt}
{\varGamma}_{\varOmega}[1]=
\begin{cases}
0& \bar{t}_{in} > \bar{t}_{out}\\
1& \bar{t}_{in} \leq \bar{t}_{out}
\end{cases},
\end{equation}
where ${\varGamma}_{\varOmega}[0]$ and ${\varGamma}_{\varOmega}[1]$ denote the first and the second bit of the mapped vector, respectively, $\bar{v}_{in}$ and $\bar{v}_{out}$ denote the average amount value of input transactions and output transactions, respectively, $\bar{t}_{in}$ and $\bar{t}_{out}$ denote the average time of input transactions and output transactions, respectively. Particularly, for the \emph{all in} patterns which has no value of $\bar{v}_{out}$ and $\bar{t}_{out}$, the corresponding ${\varGamma}_{\varOmega}$ is defined as $[0,0]$. Contrarily, ${\varGamma}_{\varOmega}$ of \emph{all out} patterns are given as $[1,1]$.

The z-score value of these patterns (Fig. \ref{z-score}) are calculated with 100 random networks for each real network. To preserve the same degree sequence and attribute distribution as the real network, these random networks are generalized by the configuration model \cite{DBLP:books/ox/Newman10} with a rearrangement of the attribute information. As displayed in Fig. \ref{z-score}, the occurring frequency of a pattern is similar in different networks since these networks are from the same domain. Pattern $a1$-$a6$ and pattern $b2$, $b4$-$b6$ are statistically significant motifs in the Bitcoin transaction network with a much greater number of occurring times in a random network. We then make use of these hybrid motifs, including temporal motifs $a1$-$a6$ and ATH motifs $b2$, $b4$-$b6$ to characterize the network features in Bitcoin transaction network. Besides, we extract some basic network features from AAIN such as in-degree, out-degree and so on. All the network features are described as follows:

\begin{table}	
	\renewcommand{\arraystretch}{1.15}	
	\caption{Average fraction of $\delta$-temporal motifs ($\delta = 3$ hours).}	
	\centering	
	\setlength{\tabcolsep}{1.15mm}{	
		\begin{tabular}{ccccccc}		
			\hline		
			\multirow{3}*{Temporal motif} & {$a1$} & {$a2$} & {$a3$} & {$a4$} & {$a5$} & {$a6$} \\
			~&{\includegraphics[scale=0.13]{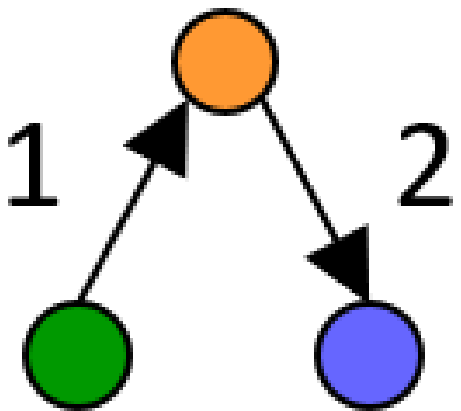}}&{\includegraphics[scale=0.13]{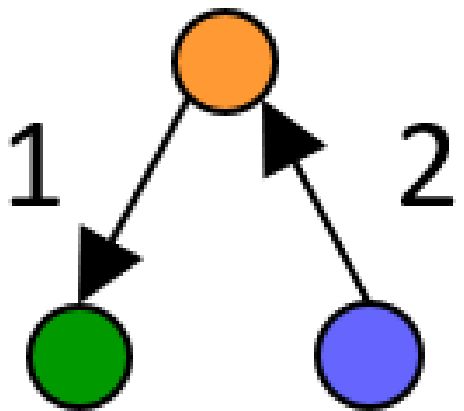}}&{\includegraphics[scale=0.13]{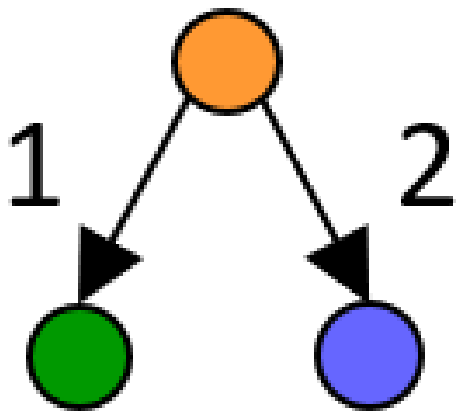}}&{\includegraphics[scale=0.13]{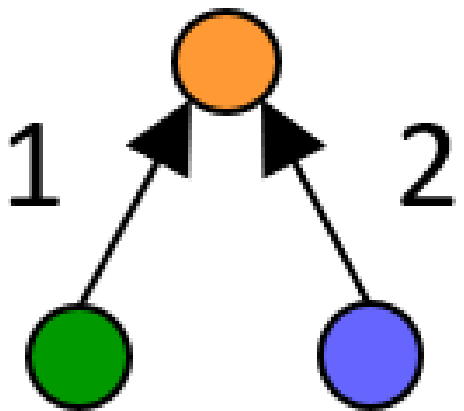}}&{\includegraphics[scale=0.13]{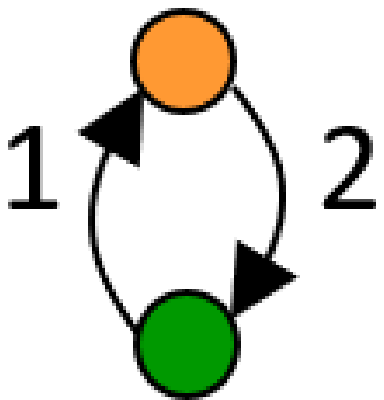}}&{\includegraphics[scale=0.13]{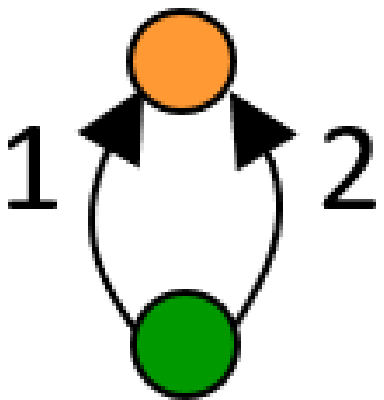}}\\		
			\hline
			\hline	
			Labeled address &  0.2552 &  0.0051 & 0.5902 & 0.1465 & 0.0000 & 0.0030\\
			\hline
			Unlabeled address &  0.2320 & 0.0576 &  0.4016 & 0.2395 & 0.0003 & 0.0690\\
			\hline
	\end{tabular}}
	\label{table_temporal_motifs}
\end{table}

\begin{table}	
	\renewcommand{\arraystretch}{1.1}	
	\caption{Average fraction of $\delta$-ATH motifs ($\delta = 3$ hours).}	
	\centering	
	\setlength{\tabcolsep}{2.7mm}{	
		\begin{tabular}{ccccc}		
			\hline		
			\multirow{3}*{ATH motif} & {$b2$} & {$b4$} & {$b5$} & {$b6$}\\
			&{\includegraphics[scale=0.13,trim=25 2 0 10]{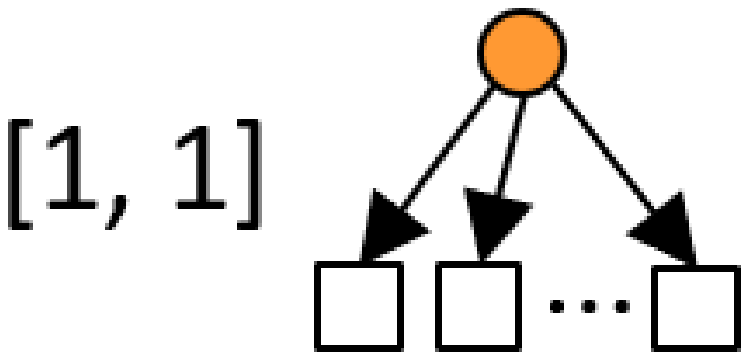}}&{\includegraphics[scale=0.13, trim=25 10 0 10]{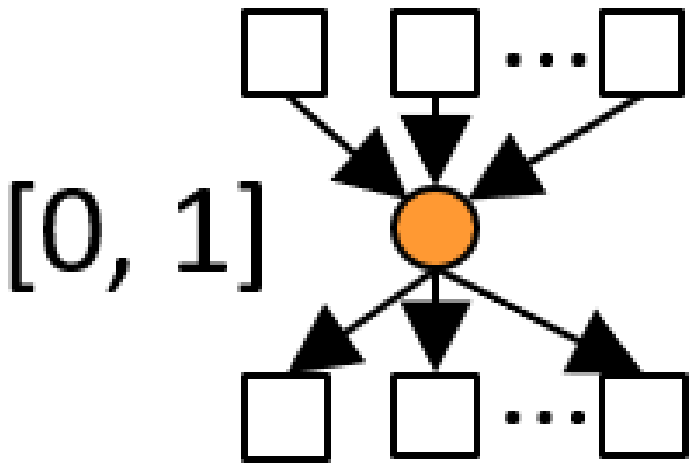}}&{\includegraphics[scale=0.13, trim=25 10 0 10]{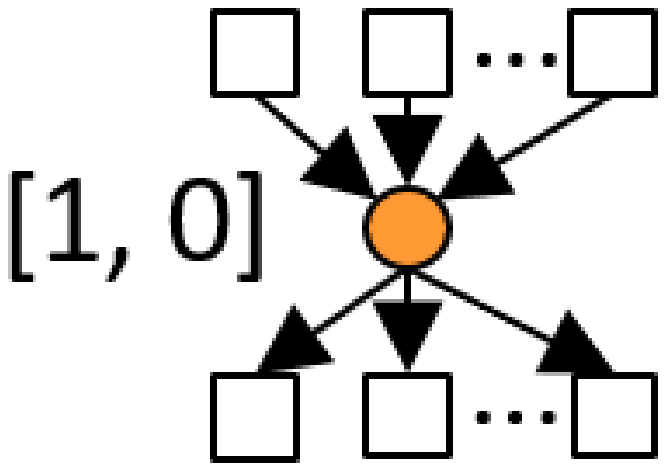}}&{\includegraphics[scale=0.13, trim=25 10 0 10]{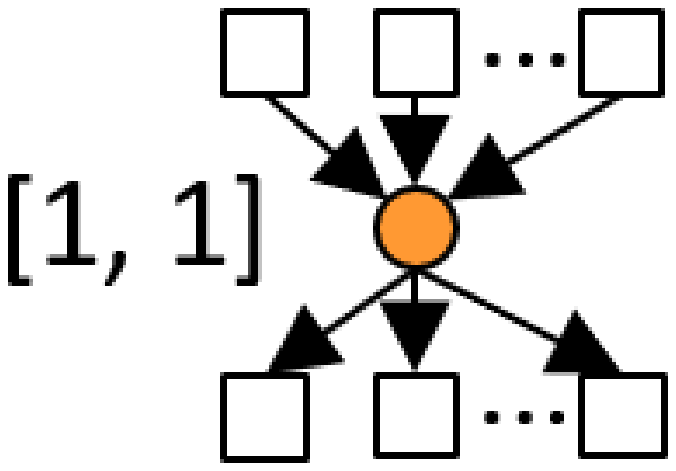}}\\
			\hline
			\hline
			Labeled address & {0.4957} & {0.4916} & {0.0057} & {0.0069}\\
			\hline
			Unlabeled address & {0.6557} & {0.3148} & {0.0069} & {0.0225}\\
			\hline
	\end{tabular}}
	\label{table_ATH_motifs}
\end{table}

NF1-NF6: The occurring frequency proportion of each temporal motif in all considered temporal motifs in the first-order AAIN for each address.

NF7-NF10: The frequency proportion of an address being a part of each ATH motif in all considered ATH motifs.

NF11: Value of in-degree in AAIN.

NF12: Value of out-degree in AAIN.

NF13: Ratio of the in-degree to the out-degree in AAIN.

NF14: Number of unique successor addresses in AAIN.

NF15: Number of unique predecessor addresses in AAIN.

NF16: Ratio of the in-degree to the number of unique successor addresses in AAIN.

NF17: Ratio of the out-degree to the number of unique predecessor addresses in AAIN.

\begin{table*}	
	\renewcommand{\arraystretch}{1.25}	
	\begin{threeparttable}[b]
		\caption{Statistics of features (except motif features).\label{table_feature}}	
		\centering
		\setlength{\tabcolsep}{0.89mm}{
			\begin{tabular}{cccccccccccccccccccc}
				\hline
				\quad& NF11&NF12&NF13&NF14&NF15&NF16&NF17&AF1& AF2& AF3& AF4& AF5 & AF6 & TF1 &TF2 & TF3 & TF4 &TF5 & TF6\\
				\hline
				\multicolumn{20}{c}{Labeled address}\\
				\hline
				Mean&5.90&9.02&0.85&5.86&8.71&1.01&1.02&1.28&1.28&1.00&1.48&1.48&1.00&0.00&5.59e+3&10.20&5.34&11.94&6.22\\
				StdDev&25.31&23.52&2.52&25.11&15.27&0.18&0.13&2.56&2.54&0.03&7.22&7.21&0.45&
				0.27
				&2.55e+4&7.83&25.46&13.89&32.80\\
				Median&2.00&10.00&0.48&2.00&10.00&1.00&1.00&1.00&1.00&1.00&0.32&0.32&1.00&0.00&1.36e+3&7.00&2.00&8.00&2.00\\
				\hline
				\multicolumn{20}{c}{Unlabeled address}\\
				\hline
				Mean&6.86&8.28&1.30&6.23&6.63&1.12&1.29&1.79&1.87&1.02&6.79&6.80&57.50&0.10&3.80e+4&31.65&59.65&38.68&132.39\\
				StdDev&91.47&222.59&5.21&43.54&75.23&5.15&5.45&42.93&43.07&0.56&1.23e+3&1.23e+3&6.88e+4&15.63&1.10e+5&95.78&375.36&129.82&833.56\\
				Median&2.00&2.00&0.76&2.00&2.00&1.00&1.00&1.00&1.00&1.00&0.11&0.11&1.00&0.00&3.31e+4&3.00&2.00&3.00&2.00\\
				\hline
		\end{tabular}}
	\end{threeparttable}
\end{table*}

Tables \ref{table_temporal_motifs} and \ref{table_ATH_motifs} describe the average value for NF1-NF10 in labeled addresses and unlabeled addresses with $\delta = 3$ hours, respectively. The selected time window will be explained in Section~\ref{TF}. And by combining with the statistics of NF11-NF17 in Table \ref{table_feature}, we can summarize several findings from network features as follows:

\begin{itemize}
	\item \textbf{Finding 1.} The average fraction of $a1$ pattern is much higher than the fraction of $a2$ pattern, and the fraction of $b4$ pattern far outstrips the other kinds of \emph{in and out} motifs $b5$ and $b6$. Besides, this kind of difference is more significant for labeled addresses. In other words, mixing services are more in line with the transaction pattern of receiving money firstly and sending money out latter with a balance not less than 0.
	
	\item \textbf{Finding 2.} Based on the results of $a5$ and $a6$ patterns, we can conclude that non-mixing service entity may reuse some addresses in a short time while this situation seldom happens for mixing services. Besides, the statistics of NF16 and NF17 also indicates the same finding.
	
	\item \textbf{Finding 3.} The prevalence of $a3$ pattern is due to the change addresses generated to receive the change. Besides, from the fraction of $a3$ as well as the statistics of NF11-NF13, we can see that mixing services prefer dispersing the tainted Bitcoins to others, which is a usually adopted strategy for Bitcoin mixing. Some analytic companies apply some taint analysis techniques \cite{MoserBB14Towards}, which can predict a risk score for addresses and blacklist the high-risk tainted coin possessors, to track these patterns and avoid buying these tainted coins.
\end{itemize}

\subsection{Account Features}
The state and activeness of an address, in many cases, may reflect which category the address belongs to, and thus we introduce account features to describe the state and activeness of an address. For example, addresses belonging to Bitcoin exchanges usually have a higher trade frequency for a great many of businesses, while the trade frequency of many ordinary users are relatively much lower. The extracted account features for each address, referred to as AFs, are detailed as follows:

AF1: Number of input transactions in the snapshot.

AF2: Number of output transactions in the snapshot.

AF3: Number ratio of the input transactions to the output transactions.

AF4: Total amount of input transactions in the snapshot.

AF5: Total amount of output transactions in the snapshot.

AF6: Amount ratio of the total input transactions to the total output transactions.

From the statistics of AF1-AF6 in Table \ref{table_feature}, some notable results can be obtained as follows:

\begin{itemize}
	\item \textbf{Finding 4.} There exists a large variety among the unlabeled addresses in terms of account features because there exist multiple types of unlabeled addresses. However, the difference of account features between labeled addresses is relatively smaller, as illustrated by a relatively low value of standard deviation.
	\item \textbf{Finding 5.} According to the results in terms of AF6, the amount value of output transactions usually equals to that of input transactions for labeled addresses, while the unlabeled addresses keep a positive net income on average, which can fully illustrate that addresses belonging to mixing services act like intermediaries by sending out what they have received.
\end{itemize}

\subsection{Transaction Features} \label{TF}
Next, the transaction behaviors of addresses in the mixing process are measured by transaction features. Since the relationship between senders and recipients of a mixing process would be obviously detected if the mixing service directly sends out an approximate equal amount to its recipients in the following blocks, mixing services may use many addresses acting like ``intermediary'' addresses (e.g. hubs) to participate in the process of fund splitting and integrating~\cite{M2014An}. After spread by a large number of intermediary addresses over a period of time, Bitcoins from the transaction sources are finally sent to the corresponding recipients.

Here we introduce the concept of \textbf{transaction cycle}, consisting of an ordered pair of continuous input and output streams, to describe the process of money flowing through an intermediary address for Bitcoins enrolled in mixing services. Fig. \ref{fig_transaction_cycle_example} displays ten continuous transactions of an address belonging to Bitcoin Fog. The x-axis in this figure represents the time line, while the y-axis represents how much the address received and sent. These ten transactions are distributed in three transaction cycles, and during each cycle, the address finally sent out what it had received with an increased balance value 0. We observe that many labeled addresses have similar transaction behavior like this, and suppose that this behavior is associated with the nature of being an intermediary. Several transaction features (referred as TFs) extracted to describe the transaction behaviors of an address are as follows:

\begin{figure}[htbp]
	\centerline{\includegraphics[scale=0.155,trim=20 20 20 20]{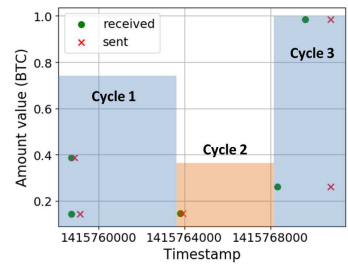}}
	\caption{A transaction cycle is composed of a continuous input stream and a continuous output stream. Three transaction cycles of labeled address ``1NsNkSxyYjB9o3QkPT2RjTXST4nGRtfMzS" are shown in this figure.}
	\label{fig_transaction_cycle_example}
\end{figure}

\begin{figure}[htbp]
	\centerline{\includegraphics[scale=0.23,trim=10 20 20 40]{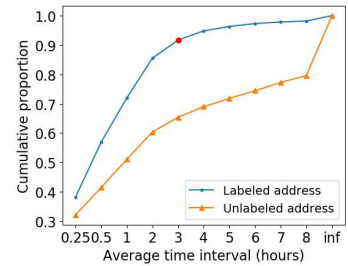}}
	\caption{The cumulative proportion of average time $\bar{T}$.}
	\label{fig_cumulative_proportion_time}
\end{figure}

TF1: Standard deviation of the increased balance in every transaction cycle (the expected value of increased balance for an intermediary address is 0).

TF2: Average time interval $\bar{T}$ between the first input transaction and the last output transaction in each cycle.

TF3: Average number of addresses that jointly participate as the inputs of a transaction.
	
TF4: Average number of addresses that jointly participate as the outputs of a transaction.

TF5: Total number of unique addresses that jointly participate as the inputs.

TF6: Total number of unique addresses that jointly participate as the outputs.

We obtain some observations based on the statistics of TF1-TF6 in Table \ref{table_feature}.
\begin{itemize}
	\item \textbf{Finding 6.} Combining with the cumulative proportion line chart of $\bar{T}$ shown in Fig. \ref{fig_cumulative_proportion_time}, we can observe that the average time interval of transaction cycles of mixing services is mostly within 3 hours, while the transaction cycle duration of an unlabeled address does not have such an obvious pattern. One possible explanation of this phenomenon is that mixing services are designed to process Bitcoins within a specific time as they are user-oriented services. As a result, we preliminarily set $\delta=3$ hours as the time window of motifs when conducting descriptive statistics and mixing detection.
	\item \textbf{Finding 7.} The TF1 results of addresses belonging to mixing services are closer to 0, indicating that the balance of these addresses over each transaction cycle is closer to 0. Besides, according to the statistics of TF3-TF6, these addresses have more co-input addresses than co-output addresses, which may be explained by that the coins of an address belonging to a mixing service are often obfuscated with coins of other addresses.
\end{itemize}


\section{Detection Model}\label{DM}
\begin{table*}
	\renewcommand{\arraystretch}{1.2}
	\begin{threeparttable}[b]	
		\caption{\label{table3}Performance comparison of different methods (with standard deviation).}	
		\centering	
		\setlength{\tabcolsep}{2mm}{	
			\begin{tabular}{c|c|c|c|c|c|c|c|c}
				\hline
				Dataset&Metric&OCSVM&IF&LR&DT&IS1$^1$&IS2$^2$&Our method\\
				\hline	
				\hline	
				\multirow{3}*{2014} & TPR & 0.8986$\pm$0.0080 & 0.8991$\pm$0.0084 & 0.1681$\pm$0.0107 & 0.7052$\pm$0.0162 & 0.8265$\pm$0.0061 & 0.8265 $\pm$0.0061& \textbf{0.9165$\pm$0.0060}\\
				\cline{2-9}
				& FPR & 0.2026$\pm$0.0064 & 0.1285$\pm$0.0139 & 0.0$\pm$0.0$^*$ & 0.0$\pm$0.0$^*$ & 0.0406$\pm$0.0003 & 0.0363$\pm$0.0003 & \textbf{0.0334$\pm$0.0010}\\
				\cline{2-9}
				& G-Mean & 0.8465$\pm$0.0033 & 0.8851$\pm$0.0068 & 0.4098$\pm$0.0130 & 0.8397$\pm$0.0097 & 0.8905$\pm$0.0033 & 0.8924$\pm$0.0033 & \textbf{0.9412$\pm$0.0029}\\
				\hline	
				\multirow{3}*{2015} & TPR & 0.8972$\pm$0.0105 & 0.8996$\pm$0.0106 & 0.0598$\pm$0.0096 & 0.6210$\pm$0.0245 & 0.7832$\pm$0.0112 & 0.7832$\pm$0.0112 & \textbf{0.9149$\pm$0.0081}\\
				\cline{2-9}
				& FPR &	0.1900$\pm$0.0124 & 0.1598$\pm$0.0147 & 0.0$\pm$0.0$^*$ & 0.0$\pm$0.0$^*$ & 0.0438$\pm$0.0003 & 0.0448$\pm$0.0003 & \textbf{0.0379$\pm$0.0016}\\
				\cline{2-9}
				& G-Mean & 0.8524$\pm$0.0047 & 0.8693$\pm$0.0075 & 0.2437$\pm$0.0197 & 0.7879$\pm$0.0156 & 0.8654$\pm$0.0062 & 0.8650$\pm$0.0062 & \textbf{0.9382$\pm$0.0038}\\
				\hline	
				\multirow{3}*{2016} & TPR & 0.8953$\pm$0.0105 & 0.9005$\pm$0.0105 & 0.0004$\pm$0.0005 & 0.3916$\pm$0.0388 & \textbf{0.9388$\pm$0.0061} & \textbf{0.9388$\pm$0.0061} & 0.9318$\pm$0.0066\\
				\cline{2-9}
				& FPR & 0.1652$\pm$0.0106 & 0.2224$\pm$0.0273 & 0.0$\pm$0.0$^*$ & 0.0$\pm$0.0$^*$ & 0.0591$\pm$0.0004 & 0.0586$\pm$0.0003 & \textbf{0.0356$\pm$0.0010}\\
				\cline{2-9}
				& G-Mean & 0.8645$\pm$0.0051 & 0.8366$\pm$0.0135 & 0.0115$\pm$0.0150 & 0.6250$\pm$0.0314 & 0.9398$\pm$0.0031 & 0.9400$\pm$0.0031 & \textbf{0.9479$\pm$0.0031}\\
				\hline	
		\end{tabular}}
		\begin{tablenotes}
			\item[1,2] IS1 and IS2 use two different inter links counting function, namely relative inter links and total inter links respectively \cite{Prado2017Discovering}.
			\item[*] The marked FPRs imply that there exist overfitting problems in LR and DT as the positive instances are more likely to be predicted as negative instances. 
		\end{tablenotes}
	\end{threeparttable}
\end{table*}

In the mixing detection task considered here, we only access a small number of verified labeled addresses belonging to mixing services while the rest addresses are unlabeled. This problem of extreme class imbalance may greatly hinder the performance of supervised classification. To deal with this problem, we develop a Positive and Unlabeled (PU) learning model with a two-stage strategy. The first stage is to select out the reliable negative instances from the unlabeled instances (unlabeled addresses) in the training set, and the second stage is to train a classifier with the positive instances (labeled addresses) as well as the reliable negative instances in the training set.

\textbf{In stage one}, according to the spy technique proposed in \cite{liu2003building}, we sample a set of spy instances from the positive instances with a default sample rate 15\%. The rest of the positive instances are set with label 1, while the spy instances as well as the unlabeled instances are set with label -1, and then they are used to train a classifier for selecting out the reliable negative instances. Here we employ the widely considered logistic regression as the classifier. Since the spy instances are actually positive instances, the probability of a spy being predicted as a positive instance would be usually higher than that of a negative instance. Therefore, we can select a threshold $\theta$ based on the prediction probabilities of the spy instances, and the reliable negative instances are selected out from the unlabeled instances if their probability of being predicted as a positive instance is lower than $\theta$. The threshold $\theta$ is selected as the value that can maximize the increment difference between the cumulative proportion of unlabeled instances and spy instances under a minute increment $\Delta p$, and it can be calculated by:
\begin{equation}\label{formula3}
\theta = {\arg\max}_{p\in[0+\Delta p,1]}(\Delta F_U(p)-\Delta F_S(p)).
\end{equation}
For each instance $i$ in the instance set, its probability of being predicted as a positive instance is denoted as $x_i$ and stored in a set $X$, namely $x_i \in X$. For the set $X$, its cumulative distribution function $F(\cdot)$ is given by
$F_X(p)=P\{X \leqslant p\}$, where $P\{X \leqslant p\}$ represents the probability that a value in $X$ is lower than or equal to a value $p$. Then the increment of $F_X(p)$ under a minute increment $\Delta p$ ($\Delta p=0.005$ in our model) is denoted as: $\Delta F_X(p)=F_X(p)-F_X(p-\Delta p)$. We use $S$ and $U$ to denote a set storing the prediction probabilities of spy instances and unlabeled instances respectively, so that $\Delta F_S(p)$ and $\Delta F_U(p)$ represent the increment of $F_S(p)$ and the increment of $F_U(p)$ under the increment $\Delta p$ respectively.

\textbf{In stage two}, with the consideration that the number of positive instances and that of reliable negative instances may be imbalanced, we set different penalty weights for different kinds of instances in the loss function. The following objective function should be minimized.
\begin{equation}
C_+\sum_{y_i=1}{l(y_i,f(\bm{x_i}))}+C_-\sum_{y_i=-1}{l(y_i,f(\bm{x_i}))}+\lambda R(\bm{w}),
\end{equation}
where $C_+$ and $C_-$ denote the penalty coefficients of positive and reliable negative instances, respectively, $l(y_i,f(\bm{x_i}))$ is the loss term, $R(\bm{w})$ is the regularization term and $\lambda$ is the regularization coefficient. In this work, we apply a biased logistic regression classifier so that the loss term is set to be log loss and the regularization term is set to be $L2$-norm. Besides, $C_+$ and $C_-$ are inversely proportional to the number of positive and reliable negative instances in our settings.

Finally, we choose a probability threshold $\varepsilon$ and make a decision according to the prediction probability of each unlabeled address. An unlabeled address is detected as an address associated with mixing services when its probability of being predicted as a positive instance is greater than $\varepsilon$.
\section{Experimental Results}\label{ER}

In this section, we conduct a comprehensive evaluation on the proposed detection framework for Bitcoin mixing services. Firstly, we describe our experimental settings. Secondly, we present the experimental results of the proposed method in comparison with several baseline methods. After that, the effects of motif-based features and other basic features are compared and summarized. Next, we demonstrate the robustness of our framework via a parameter sensitivity analysis. Finally, since our experiments are conducted on three transaction snapshots during 2014-2016, we discuss about how these data are relevant for current transactions and addresses.

\subsection{Experimental Settings}\label{ES}
We initialize the time window $\delta=3$ hours and the probability threshold $\varepsilon=0.6$. All the reported results are averaged over 100 independent experiments with the standardized features as the model inputs.

\textbf{Datasets.} As mentioned in Section \ref{DC}, we obtain three datasets with transaction data from a Bitcoin client as well as labels from WalletExplorer.
Before training the model, we filter out the addresses with either only input transactions or output transactions. By applying this simple rule, 131 labeled addresses and 1,635,904 unlabeled addresses are filtered from the three datasets, occupying 0.9\% and 22.2\% of their corresponding class, respectively. This operation is based on the following considerations. On the one hand, mixing services serve as intermediaries in obfuscating the transactions so that addresses with either only input transactions or output transactions do not satisfy this obvious feature. On the other hand, 96.2\% of the filtered labeled addresses have only one transaction record, which are not suitable for feature learning. Besides, the timestamp of all the transactions related to the filtered labeled addresses included in our datasets are close to the time boundaries of the snapshot datasets. Thus there may be some extra transactions not being captured in our three snapshots.
We then divide each dataset into training set and testing set as follows, and for each dateset, we train a model with the training set and verify the model with the testing set.
\begin{enumerate}[(1)]
	\item Training set: For stage one, we select 70\% unlabeled addresses and 70\% labeled addresses to form the training set, and then we can obtain some reliable negative instances. For stage two, the training set is made up of 70\% reliable negative instances as well as the labeled addresses used in stage one.
	\item Testing set: The testing set is formed by the remaining 30\% reliable negative instances and 30\% labeled addresses to evaluate our model.
\end{enumerate}

\textbf{Evaluation metrics.}
In this work, we evaluate the performance of our model in terms of true positive rate (TPR), false positive rate (FPR) and the Geometric Mean (G-Mean). G-Mean was suggested in \cite{kubat1997addressing} and has been widely used as a comprehensive metric in evaluating classification performances on imbalanced datasets\cite{tang2008svms}\cite{tang2016bayesian}. Taking both the accuracy of positive instances and negative instances into account, G-Mean is defined as follows:
\begin{equation}
{\rm G{-}Mean}= \sqrt{{\rm TPR} \times (1-{\rm FPR})}.
\end{equation}

\subsection{Method Comparison}\label{DR}

Our model is based on PU learning with a two-stage strategy, which is actually a semi-supervised learning method. To evaluate the effectiveness of PU learning in our scenario, we compare our model with several baseline methods including one-class support vector machine (OCSVM), isolation forest (IF) \cite{liu2008isolation}, logistic regression (LR), decision tree (DT) and InterScore (IS) \cite{Prado2017Discovering}. Among them, OCSVM and IF are two unsupervised anomaly detection method, LR and DT are two widely used supervised classifiers. IS is a Bitcoin mixing detection method which can detect mixing service entities containing multiple addresses with community anomaly detection, as addresses belonging to these entities usually have more inter-community connections than other addresses. Since here we focus on the problem of detecting addresses of mixing services, we consider the label of an address is equal to the label of its entity when implementing IS.

Table \ref{table3} compares the performance of our method with the baseline methods. Specifically, the proportion of outliers in the datasets is set to be 10\% when fitting OCSVM and IF.  According to Table \ref{table3}, we have the following observations:

\begin{enumerate}[(1)]
	\item The unsupervised anomaly detection methods (i.e., OCSVM, IF and IS) can discover most of the positive instances, however, they have a higher false positive rate than other methods. In particular, since IS only captures one important topology feature of being an intermediary in user transactions, it is in lack of generalization so that its performance is significantly differentiated in different datasets. 
	\item The two supervised methods including LR and DT lead to the problem of overfitting and relatively poor performance. There exist two possible reasons for this result, one reason is that the extreme class imbalance hinders the performance of supervised classification, and the other reason is that these two methods treat all unlabeled addresses as negative instances, which may induce noises to the datasets.

	\item By selecting reliable negative instances from unlabeled instances first and then apply a supervised method, the proposed strategy can improve the detection rate of positive instances compared with directly applying supervised approaches, and obtain the best results in terms of G-Mean.
\end{enumerate}

These observations show that the PU learning framework performs better on Bitcoin mixing detection with a high true positive rate exceeding 91\% and a low false positive rate below 4\% on extremely imbalanced datasets. 

\begin{table*}
	\renewcommand{\arraystretch}{1.15}	
	\begin{threeparttable}[b]
		\caption{\label{table5}Performance comparison of different features (with standard deviation).}	
		\centering	
		\setlength{\tabcolsep}{1.4mm}{	
			\begin{tabular}{c|c|c|c|c|c|c|c|c}
				\hline
				\multirow{2}*{Dataset}&\multirow{2}*{Metric}&\multirow{2}*{Basic features}&\multirow{2}*{Temporal motifs} &\multirow{2}*{ATH motifs} & \multirow{2}*{Hybrid motifs$^*$}& Basic features \&&  Basic features \&& Basic features \& \\
				&&&&&& Temporal motifs & ATH motifs& Hybrid motifs$^*$\\		
				\hline
				\hline
				\multirow{3}*{2014} & TPR & 0.8744$\pm$0.0145 & 0.8728$\pm$0.0070 & 0.7059$\pm$0.0111 & 0.8912$\pm$0.0064 & 0.9032$\pm$0.0064 & 0.8797$\pm$0.0089 & \textbf{0.9165$\pm$0.0060}\\
				\cline{2-9}
				& FPR & 0.1779$\pm$0.0128 & 0.0455$\pm$0.0009 & 0.1508$\pm$0.0013 & \textbf{0.0318$\pm$0.0007} & 0.0362$\pm$0.0014 & 0.1350$\pm$0.0091 & 0.0334$\pm$0.0010\\
				\cline{2-9}
				& G-Mean & 0.8479$\pm$0.0120 & 0.9127$\pm$0.0036 & 0.7742$\pm$0.0059 & 0.9289$\pm$0.0032 & 0.9330$\pm$0.0033 & 0.8723$\pm$0.0075 & \textbf{0.9412$\pm$0.0029}\\
				\hline	
				\multirow{3}*{2015} & TPR & 0.8146$\pm$0.0115 & 0.8453$\pm$0.0098 & 0.8426$\pm$0.0100 & 0.8823$\pm$0.0088 & 0.8864$\pm$0.0092 & 0.8543$\pm$0.0095 & \textbf{0.9149$\pm$0.0081}\\
				\cline{2-9}
				& FPR &	0.1388$\pm$0.0038 & 0.1423$\pm$0.0024 & 0.0716$\pm$0.0009 & 0.0667$\pm$0.0020 & 0.0878$\pm$0.0079 & 0.0852$\pm$0.0018 & \textbf{0.0379$\pm$0.0016}\\
				\cline{2-9}
				& G-Mean & 0.8376$\pm$0.0064 & 0.8515$\pm$0.0043 & 0.8845$\pm$0.0051 & 0.9074$\pm$0.0041 & 0.8992$\pm$0.0065 & 0.8840$\pm$0.0048 & \textbf{0.9382$\pm$0.0038}\\
				\hline	
				\multirow{3}*{2016} & TPR & 0.6442$\pm$0.0317 & 0.9271$\pm$0.0073 & 0.6639$\pm$0.0145 & 0.9123$\pm$0.0071 & \textbf{0.9335$\pm$0.0067} & 0.8150$\pm$0.0112 & 0.9318$\pm$0.0066\\
				\cline{2-9}
				& FPR & 0.3812$\pm$0.0077 & 0.0584$\pm$0.0012 & 0.3154$\pm$0.0043 & 0.0356$\pm$0.0011 & 0.0508$\pm$0.0011 & 0.1995$\pm$0.0047 & \textbf{0.0356$\pm$0.0010}\\
				\cline{2-9}
				& G-Mean & 0.6311$\pm$0.0129 & 0.9343$\pm$0.0035 & 0.6741$\pm$0.0061 & 0.9380$\pm$0.0034 & 0.9413$\pm$0.0031 & 0.8077$\pm$0.0047 & \textbf{0.9479$\pm$0.0031}\\
				\hline	
		\end{tabular}}
		\begin{tablenotes}
			\item[*] Hybrid motifs are a combination of Temporal and ATH motifs. 
		\end{tablenotes}
	\end{threeparttable}
\end{table*}

\begin{figure*}
	\centering
	\subfigure[TPR measured against different $\delta$.]{
		\label{fig8a}
		\includegraphics[scale=0.22,trim=10 20 10 30]{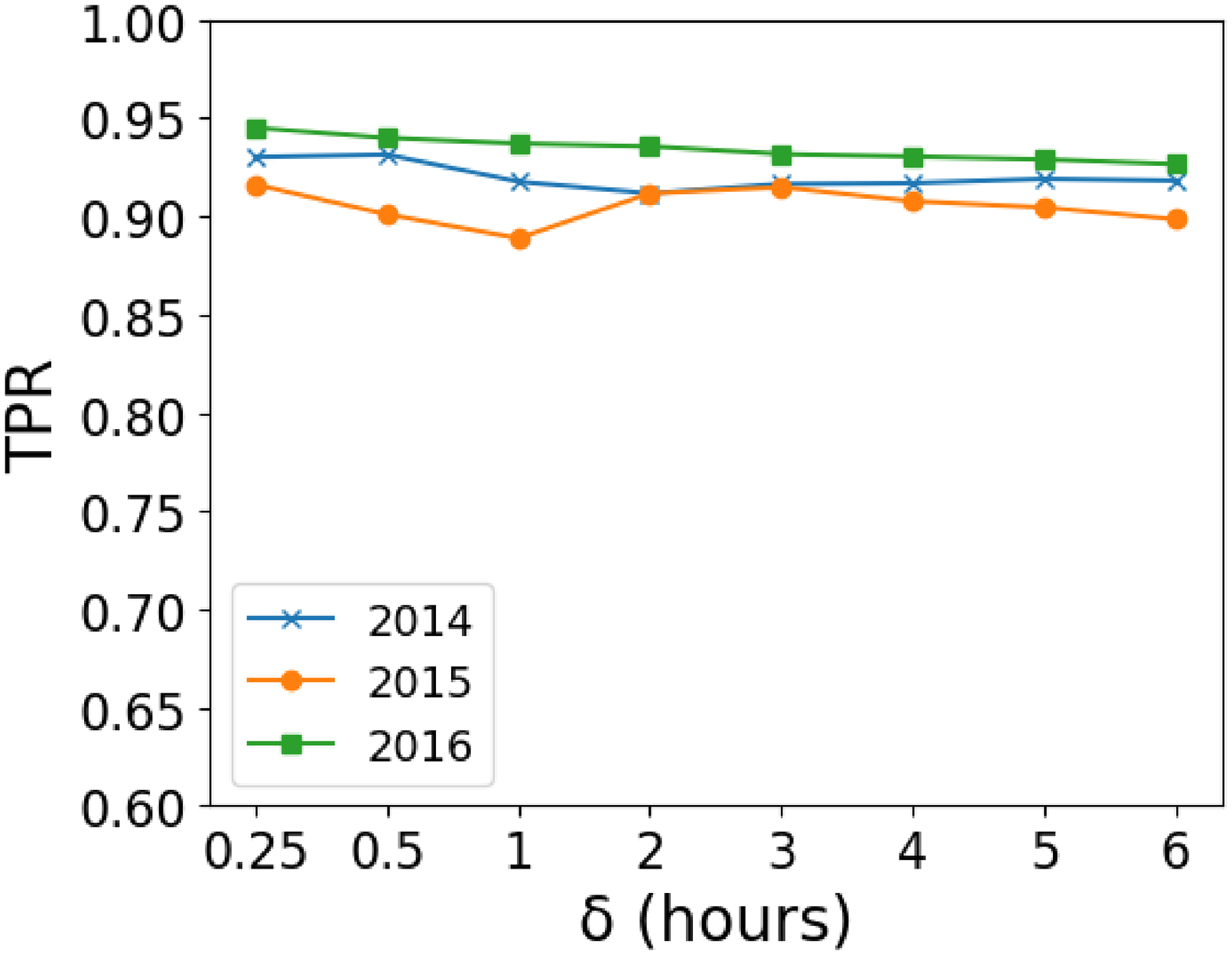}}
	\subfigure[FPR measured against different $\delta$.]{	
		\label{fig8b}
		\includegraphics[scale=0.22,trim=10 20 10 30]{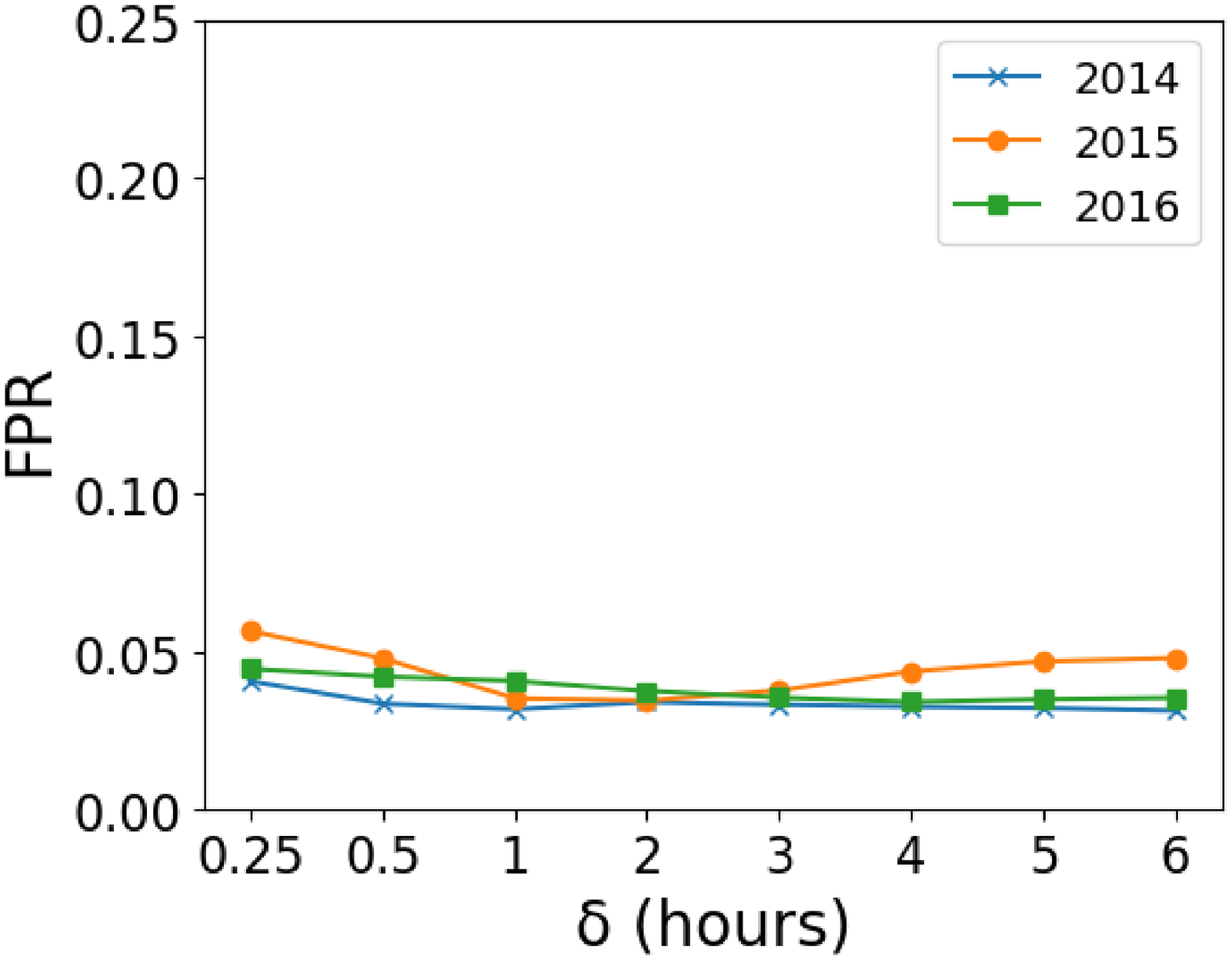}}
	\subfigure[G-Mean measured against different $\delta$.]{	
		\label{fig8c}
		\includegraphics[scale=0.22,trim=10 20 20 30]{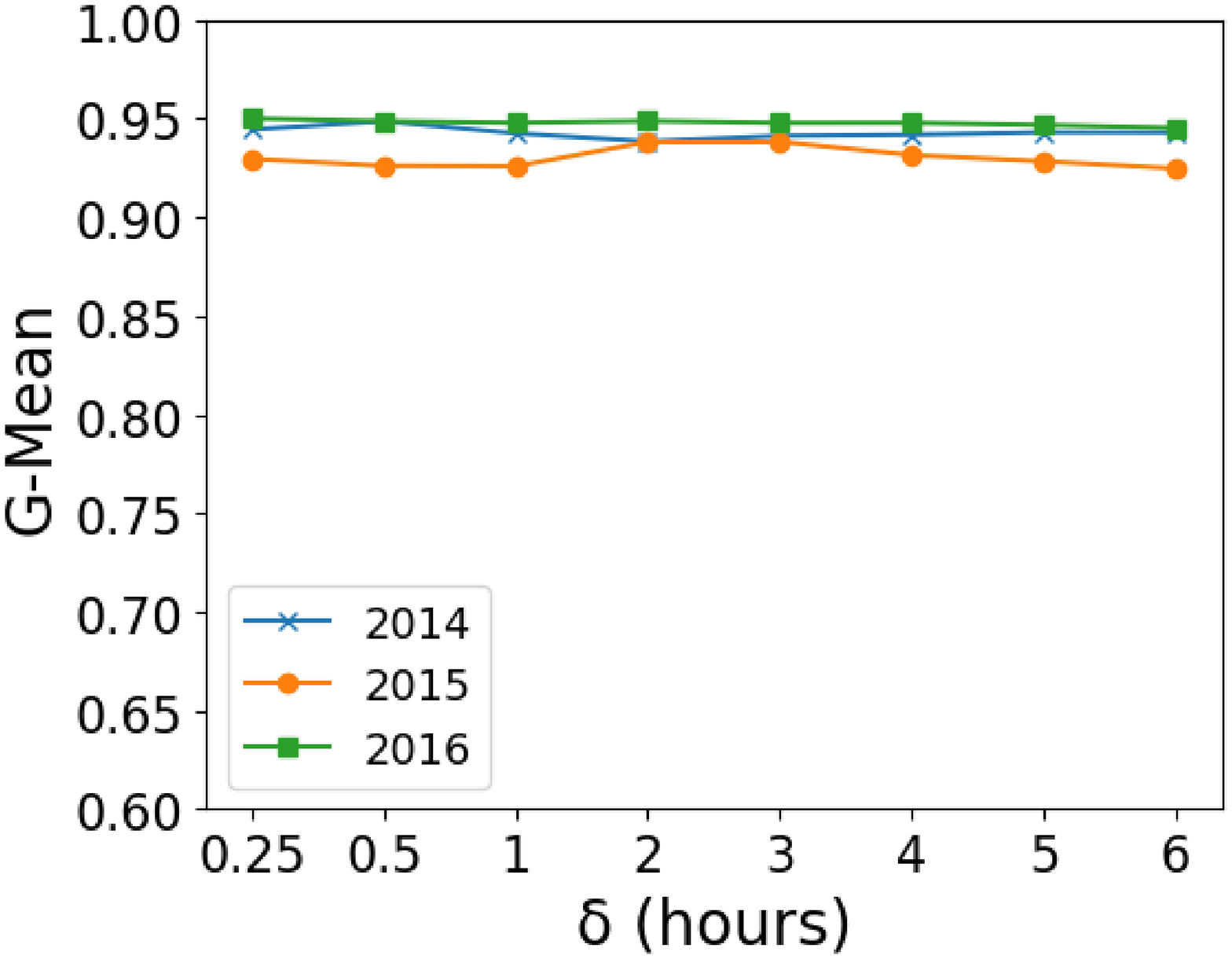}}
	\caption{Patameter analysis of time window $\delta$. (a), (b), (c) and (d) show the results of TPR, FPR and G-Mean measured against different $\delta$ respectively.}
	\label{fig_delta}
\end{figure*}

\subsection{Feature Performance Comparison}
{To verify the effectiveness of the proposed motif features, we divide all the features given in Section IV into basic features (features except motifs) and motifs (including temporal and ATH motifs) to train the classifier and further evaluate the importance of motifs in the detection. A detailed comparison is given in Table \ref{table5} and can be summarized as follows:
\begin{enumerate}[(1)]
	\item The detection performance is relatively poor when we only use the basic features. While network motifs, which can reveal the higher-order features in complex network, achieve decent performance in mixing detection.
	\item Each evaluation metric can be significantly improved for almost all cases when combining hybrid motifs with basic features, which demonstrates that hybrid motifs play an indispensable role in the task of Bitcoin mixing detection.
\end{enumerate}

Additionally, since LR has a good explainability in its model weights, which can reflect the influence degree of different features to the detection result, we analyze the impact of the features according to their absolute value of weight averaged over the three datasets in the 100 independent experiments. We find that the top 10 important features are TF3, AF5, AF4, NF17, NF11, NF16, NF12, NF14, TF6 and TF1, illustrating that the basic features also play an important role in the detection process.

\begin{table}
	\renewcommand{\arraystretch}{1.15}	
	\caption{Patameter analysis of probability threshold $\varepsilon$.}
	\centering	
	\setlength{\tabcolsep}{1.8mm}{	
		\begin{tabular}{c|c|c|c|c|c|c}
			\hline
			Dataset&Metric&0.5&0.6&0.7&0.8&0.9\\
			\hline
			\hline	
			\multirow{2}*{2014} & TPR &\textbf{0.9318}&0.9165&0.8962&0.8653&0.8016\\
			\cline{2-7}
			& FPR &0.0535&0.0334&0.0191&0.0093&\textbf{0.0028}\\
			\hline	
			\multirow{2}*{2015} & TPR &\textbf{0.9339}& 0.9149 &0.8863&0.8420&0.7652\\
			\cline{2-7}
			& FPR &	0.0686 & 0.0379 &0.0182&0.0083&\textbf{0.0032}\\
			\hline	
			\multirow{2}*{2016} & TPR &\textbf{0.9438}&0.9318&0.9184&0.8990&0.8510\\
			\cline{2-7}
			& FPR &0.0502&0.0356&0.0221&0.0114&\textbf{0.0037}\\
			\hline	
	\end{tabular}}
	\label{table6}
\end{table}

\subsection{Parameter Sensitivity Analysis}
Next, we provide a sensitivity analysis for the time window parameter $\delta$ and the probability threshold $\varepsilon$ to understand their impacts on the performance of the proposed model.

\begin{table}
	\renewcommand{\arraystretch}{1.15}
	\begin{threeparttable}[b]
	\caption{\label{2020result}Performance comparison of different methods on current transaction data (With standard deviation).}
	\centering	
	\setlength{\tabcolsep}{1.3mm}{	
		\begin{tabular}{c|c|c|c|c}
			\hline
			Dataset&Method&TPR&FPR&G-Mean\\
			\hline
			\hline	
			\multirow{7}*{2020$^{1}$} &OCSVM &0.6474$\pm$0.0974&\textbf{0.0572$\pm$0.0162}&0.7788$\pm$0.0567\\
			\cline{2-5}
			&IF &\textbf{0.8537$\pm$0.0774}&0.5431$\pm$0.0539&0.6220$\pm$0.0350\\
			\cline{2-5}	
			&LR &0.0$\pm$0.0&0.0$\pm$0.0$^*$&0.0$\pm$0.0\\
			\cline{2-5}
			&DT &0.0685$\pm$0.0599&0.0$\pm$0.0$^*$&0.2236$\pm$0.1368\\
			\cline{2-5}
			&IS1 &0.8189$\pm$0.0629&0.4121$\pm$0.0007&0.6933$\pm$0.0268\\
			\cline{2-5}	
			&IS2 &0.8189$\pm$0.0629&0.4006$\pm$0.0008&0.7001$\pm$0.0270\\
			\cline{2-5}
			&Our method &0.7874$\pm$0.0671&0.0991$\pm$0.0104&\textbf{0.8413$\pm$0.0327}\\
			\hline			
	\end{tabular}}
	\begin{tablenotes}
		\item[1] This dataset contains 1,500,000 consecutive transactions since Mar. 25, 2020.
		\item[*] The marked FPRs imply that there exist overfitting problems in LR and DT as the positive instances are more likely to be predicted as negative instances. 
	\end{tablenotes}
	\end{threeparttable}
\end{table}

Fig.~\ref{fig_delta} shows the results in terms of TPR, FPR and G-Mean of our model versus time window $\delta \in \{0.25, 0.5, 1, 2, ..., 6\}$ hours. We can observe that the curves of the metrics are generally stable, which illustrates that our model can steadily obtain relative good results under different settings of parameter $\delta$ in the testing domain. 


We also provide the TPR and FPR results of our model under different probability threshold $\varepsilon$ for voting in Table \ref{table6}. We can observe that the lower $\varepsilon$ is, the higher TPR is. While for FPR, it becomes lower with a larger $\varepsilon$. For practical applications, we can choose an appropriate threshold according to our specific requirement of pursuing a higher TPR or ensuring a lower FPR.

\subsection{Discussion}
Recent years, some new techniques like Segregated Witness (SegWit) \cite{lombrozo2015bip141} and Lightning Network (LN) \cite{poon2016bitcoin} have been developed in the Bitcoin community, bringing some new changes to Bitcoin transactions and addresses. The SegWit soft-fork accepted in August, 2017 is a solution for the transaction malleability problem \cite{DBLP:conf/esorics/DeckerW14} by redesigning the transaction structure and segregating the witness data so that modifications to the witness would not change the transaction hash. In addition, the use of SegWit can reduce the size of a transaction and increase the number of transactions contained in a block. LN is an off-chain solution to improve the scalability of Bitcoin. For two Bitcoin users, they can open a LN channel by locking some Bitcoins in a 2-of-2 multi-signature address through an on-chain transaction, and then they can trade with each other via this channel without recording in Bitcoin. Once they broadcast a commitment transaction onto the blockchain to get their respective balance from the multi-signature address, the LN channel will be closed. Different from traditional transactions in Bitcoin, the use of LN only results in two transaction records each time for the opening and closing of a channel. Besides, almost all the LN transactions are based on 2-of-2 multi-signature addresses in witness scripts after the activation of SegWit, and these addresses are native Segwit addresses started with ``bc".
	
Since WalletExplorer does not include the new emerging services after 2016, we have conducted experiments with three snapshots during 2014-2016. To justify how the data in our experiments are relevant for current transactions and addresses, we collect 1,500,000 consecutive transactions on Bitcoin since Mar. 25, 2020 and examine the performance of our model on the dataset which is referred to as 2020 dataset in Table \ref{2020result}. The active labeled addresses of mixing services crawled from WalletExplorer in this snapshot are 89 addresses belonging to Bitcoin Fog, and the number of unlabeled addresses is 1,654,175 after filtering. We also conduct a method comparison experiment on this new dataset under the same experimental settings as Section \ref{DR}. The performance comparison results displayed in Table \ref{2020result} show that our model still performs best in terms of G-mean. Yet its performance in terms TPR and FPR is slightly worse than the best method.
These results may be due to two possible reasons. One is the small amount of labels in the 2020 dataset, the other one is the introduction of some new techniques like LN. To further enhance the detection effectiveness on current transactions and addresses, the most direct method is to collect more labels via using some mixing services and then conduct analysis on them for better capturing their features. Another feasible solution is to utilize link prediction to enrich the link information of native Segwit addresses, which can help us better identify their ownership.

\section{Related Work}\label{RW}

As a new technology, blockchain has attracted intensive interests of researchers from various fields. Since the transaction data of blockchain systems are publicly accessible, they have been extensively studied to mine some network properties for transaction networks \cite{Reid2013,8575170,Ron2013
}, to cluster addresses sharing the same ownership \cite{8467371, Androulaki2013}, to discover some specific activities such as scams \cite{vasek2015there, chen2018detecting, 8668768}, attacks \cite{Tschorsch2016} and dark market trading \cite{christin2013traveling}, etc. Chen et al. conducted a graph analysis and abnormal contract detection on Ethereum with money flow graph, smart contract creation graph and smart contract invocation graph \cite{chen2018understanding}. Tam et al. proposed a Graph Convolution Network (GCN)-based embedding method to identify illicit accounts within the e-payment networks including the Ethereum transaction network \cite{Tam2019identifying}. In \cite{chen2019market}, typical abnormal transaction patterns for Bitcoin market manipulation were mined by inspecting the base networks with singular value decomposition metrics.

For the issue of money laundering detection in Bitcoin, M\"{o}ser et al. provided an inquiry into the operation models of three mixing services and tried to trace the anonymous transactions \cite{M2014An}. Weber et al. emphasized the importance of Anti-money laundering (AML) regulations in financial system, and contributed the Elliptic dataset for illicit activity detection in Bitcoin \cite{Weber2019anti}. Ranshous et al. introduced the idea of motifs in directed hypergraphs and recognized some specific laundering patterns for Bitcoin exchanges \cite{ranshous2017exchange}. Bitconeview, a visualization tool for Bitcoin, was proposed to visualize how and when an address mixes its money~\cite{Battista2015Bitconeview}. Recently, a cryptocurrency exchange platform called ShapeShift\footnote{https://classic.shapeshift.com/.} was reported to be involved in money laundering activities by moving Bitcoins to other privacy-enhancing cryptocurrencies such as Zcash \cite{kappos2018empirical} and Monero \cite{moser2018empirical}. To address this problem, Yousaf et al. proposed recognition methods for tracing cross-ledger transaction behaviors \cite{Yousaf19Tracing}.
Yet these techniques do not focus on identifying addresses enrolling in mixing. Another work shed light on the problem of Bitcoin mixing detection and tackled it as a community outlier detection problem~\cite{Prado2017Discovering}. However, this work is in lack of generalization for different mixing services and it only utilizes the topology information of transaction network. Inspired by a related study about detecting Ponzi schemes on Ethereum \cite{chen2018detecting}, in this paper, we propose features from multi-level, trying to discover the transaction patterns of mixing services for enhancement of the generalization ability.

It is worth mentioning that the network motifs we used, which are defined as the recurrent subgraph patterns of complex networks \cite{milo2002network}, play an important role in characterizing the behavior of mixing services. As the simple building blocks in complex systems, motifs have been demonstrated as a powerful tool for revealing higher-order organizations \cite{Benson2016Higher} and functional properties. Since many interactions between objects are intermittent rather than persistent, network motifs combined with temporal information were proposed to characterize dynamic homogeneous network \cite{paranjape2017motifs}, and also had an extensive version in heterogeneous information network~\cite{li2018temporal}. Recently, there are many studies utilized network motifs in blockchain transaction network mining tasks, such as price prediction \cite{akcora2018forecasting, Abay2019ChainNet}, network property analysis \cite{moreno2018mind}, exchange pattern mining \cite{ranshous2017exchange} and so on. Network attributes play important roles in network mining tasks \cite{li2018community}, nevertheless, most of these network mining studies fail to consider the rich information of network attributes when characterizing the interaction patterns with motifs.

\section{Conclusion and Future Work}\label{C}
In this work, we studied the Bitcoin mixing detection problem and conducted a systematic analysis to characterize how addresses belonging to mixing services behave in the Bitcoin transaction network. To mine the dynamic process and transaction patterns in Bitcoin more comprehensively, we employed the Bitcoin transaction records to build two temporal directed graphs including a homogeneous Address-Address Interaction Network (AAIN) and a heterogeneous Transaction-Address Interaction Network (TAIN). For TAIN, we proposed a novel concept of ATH motifs to integrate edge attribute information with higher-order structures. We developed hybrid motifs, including temporal motifs in AAIN and ATH motifs in TAIN, as the key features for mixing detection. With several designed features, we built a PU learning based detection model to handle the issue of extremely label imbalance of the mixing detection problem. Extensive experimental results on three real Bitcoin datasets demonstrated the effectiveness of our detection model. 

This work revealed some critical transaction behaviors which can distinguish the addresses belonging to mixing services, and then designed an effective method to detect these addresses. One concern is that, the mixing service providers may update their mechanisms to eliminate these typical behaviors and avoid being detected. For example, they can inject extra Bitcoins from external addresses and those injected tainted Bitcoins may sit in their addresses for a long time to fake the flow of Bitcoins, or they may increase and randomize the interval between the arrival and departure of Bitcoins, to avoid creating the discussed motifs within specific time window. Since the available data is intrinsically mostly unlabeled and our detection model is based on the prior information, these unknown complex mixing strategies may exist and may not be detected.
For future work, we will look for more adaptive strategies to defense these updated privacy-enhancing techniques, such as applying link prediction to enrich the money flow information.

\ifCLASSOPTIONcaptionsoff
  \newpage
\fi



%
\bibliography{ref}
\bibliographystyle{IEEEtran}
%








\end{document}